# Displacive model of deformation twinning in hexagonal close-packed metals. Case of the (90°, **a**) and (86°, **a**) extension twins in magnesium

Cyril Cayron

Laboratory of ThermoMechanical Metallurgy (LMTM), PX Group Chair, Ecole Polytechnique Fédérale de Lausanne (EPFL), Rue de la Maladière 71b, 2000 Neuchâtel, Switzerland

December 2016

## Abstract

A crystallographic displacive model is proposed for the extension twins in magnesium. It is based on a hard-sphere assumption previously used for martensitic transformations. The atomic displacements are established, and the homogeneous lattice distortion is analytically expressed as a continuous angular-distortive matrix that takes the usual form of shear when the distortion is complete. The calculations prove that a volume change of 3% occurs for the intermediate states and that the twinning plane, even if untilted and restored when the distortion is complete, is not fully invariant during the transient states. The crystallographic calculations also show that the (90°, **a**) twins observed in magnesium nano-pillars and the (86°, **a**) twins observed in bulk samples come from the same mechanism, the only difference being the existence of a slight obliquity angle ($\pm 3.4°$) required to reduce the strains in the latter case. Continuous features in the pole figures between the low-misoriented (86°, **a**) twin variants are expected; they are confirmed by EBSD maps acquired on a deformed magnesium single crystal. As the continuous mechanism of extension twinning is not a simple shear, a "virtual work" criterion using the value of the intermediate distortion matrix at the maximum volume change is proposed in place of the usual Schmid's law. It allows predicting the formation of extension twins for crystal orientations associated with negative Schmid factors.

**Keywords :** Extension twins; hexagonal close-packed; magnesium; Schmid factor; angular-distortive matrix; hard sphere.

## 1. Introduction

### 1.1. The choice of paradigm

It seems important to make clear the paradigm used in this work. Indeed, the manuscript in its first formulation, i.e. without the sections 1.1, 5, 6.2 and 6.3 [1], was rejected twice mainly because it does not imply "disconnections" [2]. The reviewers also estimated that the model was not realistic and was an oversimplification [3]. We hope that this preamble will help the reader to understand why we think that the mechanism of deformation twinning is not compatible with the "disconnection" theory, and why the "displacive" paradigm has been chosen. We admit that the electronic structure of the crystal is ignored in first approximation, but a simple model is not a false model; it is just a first step toward a more accurate model.

Deformation twins appear in the face centered cubic (fcc) and body centered cubic (bcc) metals deformed at high speeds or low temperatures. They are more commonly formed in hexagonal close-packed (hcp) metals because of the lowest number of slip modes. Deformation twinning share many characteristics with martensitic transformations; they are formed at very high speed (close to the speed of sound) and take the shape of lenticular plates that become highly intricate at high deformation rates: both deformation twinning and martensitic transformations belong to the wide class of displacive transformations. Deformation twinning has been mathematically treated as a *homogeneous* shear for more than one century [1]-[11]. Bevis and Crocker [8] adopted the definition "*twinning shear is any shear which restores a lattice in a new orientation*" and used it to build a generalized theory that predicts the possible twinning matrices. The more realistic ones were chosen among those with the minimum shearing magnitude. Very early, the use of shear matrices relies on the observations of planar interfaces between the parent and its twins. As explained by Christian and Mahajan [9]: "*Since a parent crystal and its twin remain in contact at the interface plane during the formation of the twin, the relation between the structures must be such that this plane is invariant in any deformation carrying one lattice into the other*". Consequently, the classical displacive "shear" theory of twinning makes a large use of shear planes and shear directions, or their conjugates. Concerning more particularly the hcp metals, fifteen deformation twinning modes could be established in titanium by Crocker and Bevis [10], and a more recent theoretical study also based on lattice correspondence resulting from shear deformation has been proposed by Niewczas to analyze the correspondence of the dominant slip modes in the parent and its twins [11]. Another point should be mentioned. The Bravais cell of a hcp crystal is composed of two atoms, roughly speaking the 8 atoms at the "corners" of the hexagonal lattice (counting for 1/8) and one atom in the position (2/3, 1/3, 1/2) "inside" the hexagonal cell. The atoms at the corners have a trajectory directly given by the homogeneous linear shear transformation, while the trajectory of the atom inside the cell is given by an affine transformation, i.e. the same linear shear distortion, but associated with a translation component called "shuffle". The homogeneous lattice distortion is in full agreement with the fact that the atoms can move collectively in a coordinated way at very high velocities.

Although the "displacive shear" theory is established for more than a century, another approach was proposed twenty years ago by Pond, Hirth, Serra, and Bacon [12][13][14]. They introduced the concept of "disconnection", derived from the "twinning dislocations" introduced by Orowan in 1954 [15], "emissary dislocations" imagined by Slesswyk in 1963 [16], and "zonal dislocations" introduced by Mendelson in 1969 [17]. Their theory, also named "topological model", supposes that the lattice is *inhomogeneously* transformed by the movement of disconnections. A disconnection is a "customized" defect; it is a mixture of a classical dislocation with its usual Burgers vectors and a "step height" introduced in order to accommodate the geometrical misfit at the interface. The disconnections are determined with the help of the dichromatic pattern i) by overlapping the parent and twinned lattices according to their orientation relationship (OR), and ii) by finding the smallest displacements vectors between the nodes of the two lattices. Disconnections are a substitute to the displacements resulting from the homogenous lattice distortion in the classical displacive theory.

The disconnection theory seems to rally an active community despite of its major weaknesses:

a) The concept of "shuffle" has strongly digressed from the one used in the classical shear theory. In the disconnection theory, the shuffles are associated with the motion of the step part of the disconnection, without clear justification of this choice and no apparent link with the initial meaning of "shuffle" because the notion of homogeneous lattice distortion was completely suppressed.

b) The disconnection theory can't explain where the dislocations/disconnections come from and how they can be produced in a very precise sequential and coordinate order. Serra and Bacon [13] considered (as we do) that the "pole mechanism" is not realistic and proposed another model where "*a matrix dislocation becomes a new source of twinning dislocations*", but after twenty years the existence of these sources remain to be proved experimentally. In addition, nothing is said about their driving forces. These forces can't be mechanical in the case of martensitic transformation induced by cooling, and for deformation twinning it is not clear why the mechanical stress would favor the propagation of these hypothetical "twinning" dislocations instead of the usual dislocations.

c) The disconnection theory is unable to account for the elastic phase transformations and for the second order phase transitions (in ceramics and ferroelectrics) because in these cases dislocations are not required to accommodate the parent/product interfaces.

d) It is mathematically possible to associate to a homogeneous displacement field by an inhomogeneous distribution of disconnections; however, the notions of lattice distortion and correspondence that specifies how the crystallographic directions are transformed are then lost. By assuming that the shear in inhomogeneous, it is nearly impossible to figure out how the atoms move. For example, the Bain correspondence of fcc-bcc martensitic transformation can't exist within the disconnection theory.

e) Any theory should be judged by its predictions. The results of the disconnection theory are often compared with High Resolution Transmission Electron Microscopy (HRTEM) images or molecular dynamic simulations, but the macroscopic features of the transformations (habit planes, variant pairing, strains accumulated around the twins etc.) are clearly beyond its scope.

f) Because heterogeneous deformation, dislocations and plasticity are the results of irreversible mechanisms, the "disconnection" theory is unable to explain the detwinning mechanisms observed in fcc [18] and hcp metals [19][20] or the reversibility of the transformations in shape memory alloys.

g) The disconnection theory is unable to explain how a deformation twins can grow at the speed of sound. How is it possible that coordinate arrays of dislocations can move at such high velocities? In the same order of ideas, why decreasing the temperature favors deformation twins if they are generated by dislocations?

By ignoring the basic principles of the classical displacive theory, the disconnection theory becomes unable to respond to these fundamental questions. To our opinion, dislocations are not the cause of the twinning distortion but its consequence; they are the defects let at the interface, and also far away from the interface in the surrounding matrix; they are created to locally accommodate the deformation caused by the lattice distortion. Moreover, we think that the concept of "moving interface" gives a false image of the mechanism. Martensite and deformation twinning is suddenly formed in a cooperative (military) way, and all the atoms move similarly relatively to their neighbors:

the interface does not move, it is the product phase that is transformed by bursts inside the surrounding matrix. The classical displacive theory uses homogeneous distortion (linear algebra) that is well adapted to describe the formation of twins or martensite that grows at high velocities as a "wave" [21][22]. For all these reasons, we have chosen to treat deformation twinning within the displacive paradigm, i.e. by using distortion matrices, correspondence matrices etc. This is the approach we have already followed in our recent developments of the crystallography of martensitic transformation in steels [23], and more generally between the fcc, bcc and hcp phases [24]. The case of mechanical twinning in fcc crystals was also treated in Ref.[24]. The main difference between our model and the classical "shear" theory is that we believe that shear matrices are inappropriate to catch the details of the distortion mechanism because the atoms would interpenetrate too much during a continuous simple shear. Therefore, we have replaced the shear matrices by the more general "angular distortive" matrices [24] that can be explicitly calculated by assuming that the atoms are hard-spheres of constant size. The displacive paradigm does not mean that the concept of disconnection is useless but, to our opinion, disconnections and disclinations should be introduced in a second stage, as consequences of the lattice distortion. The disconnections are the dislocation arrays that accommodate the translation parent/twin misfits at the interface, and the disclinations are the dislocation arrays that accommodate the rotational misfits in the parent matrix around the twin.

### 1.2. Current issues about extension twinning in magnesium

Magnesium is a very good example of hcp metal in which deformation twinning play a major role on the mechanical properties. Magnesium, thanks to its lightness, is used in some automotive parts and is considered as a good candidate for many other applications; however, it suffers from a poor ductility because its number of slip systems is low. Twinning comes as an additional deformation mechanism that improves the elongation, but in a very anisotropic way. A better understanding of deformation twinning could thus help to improve the mechanical properties and enlarge the use of magnesium alloys in industry. The main twinning mode observed in magnesium is "extension" twinning on the $\{10\bar{1}2\}$ planes. These twins are formed when the **c**-axis is close to the tensile axis. Other twinning modes can also be observed, such as the "compression" twins on the $\{10\bar{1}1\}$ planes that are created when the **c**-axis crystal is close to the compression axis, and the so-called $\{10\bar{1}1\}$-$\{10\bar{1}2\}$ "double"-twins that are supposed to the result from a double-step compression-extension mechanism. The twinning modes were historically determined from the trace of the habit planes in deformed single crystals [25], and they are now currently and automatically identified in single crystal or polycrystalline magnesium by Electron Back Scatter Diffraction (EBSD) by measuring the specific misorientations between a parent grain and the twin (see ref. [26] for example). The twin boundaries of these three twinning modes are all characterized by a rotation around the **a**-axis and differ only by their rotation angle: 86° for the $\{10\bar{1}2\}$ extension twins, 56° for the $\{10\bar{1}1\}$ contraction twins, and 38° for the $\{10\bar{1}1\}$-$\{10\bar{1}2\}$ "double-twins". Many questions concerning the twinning mechanisms remain open. The $\{10\bar{1}2\}$ extension twins and the $\{10\bar{1}1\}$ compression twins were on the list of the fifteen twin modes predicted by Crocker and Bevis [10], but not the "double-twins", and many of the predicted modes could not be observed. More surprisingly, many experimental studies report that twins can form despite low or even negative Schmid factors; this effect was called "anomalous twinning" or "non-Schmid behavior". Anomalous twinning was reported mainly for the double-twins and compression twins [25][27][28][29], and more recently for the extension twins [30]-[33]. Another unsolved question is the atomic displacements during twinning. As mentioned in the previous section, according to the classical "shear" theory, the atoms at the lattice nodes follow the homogeneous lattice shear displacements, and the other atoms shuffle. Even if it is agreed that both shear and shuffles are concomitant, the "shear" theory first establishes the lattice shear, and only then estimates the shuffles without rigorous justification. Very recently, Yu et al. [34] developed a two-step rotation-shear model of the atomic trajectories followed during extension twinning in which the shear and shuffle displacements occur simultaneously. The model we will propose can be understood as a one-step model, where the combination of rotation and shear appears naturally as a consequence of the hard-sphere assumption.

Another point is puzzling. Recently, Liu et al. [35][36] reported in-situ transmission electron microscopy (TEM) observations of extension twins in a submicron-sized single crystal magnesium pillar induced by compression along the $[1\bar{1}0]$ axis. Surprisingly, the orientation relationship between the parent crystal and its twin is a rotation of 90° around the **a**-axis, in place of the expected 86° rotation. In addition, the parent/twin interface is made of basal/prismatic terrace-like interfaces instead of a straight boundary along a $\{10\bar{1}2\}$ plane. They also observed that the interface can "propagate" reversibly without obvious shear. These results made the authors assume that the extension twins they observed were not produced by a simple shear but by a "*newly discovered deformation mode*". This mode implies a direct conversion between the basal and prismatic planes of the parent crystal and its twin resulting in a "tetragonal compression" of the lattice, which was called "unit cell reconstruction". This "reconstructive" model contradicts the disconnection theory, but is also unable to account for the displacive nature of mechanical twinning; it was received with reservations [12][37]. For example, Ostapovets *et al.* [37] gave another explanation to the experimental results. They showed that the orthorhombic deformation matrix is the stretch component that appears in the polar decomposition of the usual simple shear matrix, and they proposed that the 90° twins are "produced by an average of two conjugate simple shears". However, the traces of the two compensating shearing planes could not be observed by Liu *et al.* in the submicron-sized pillars. The controversy that then emerged between the partisans of the disconnection theory and the reconstructive model is explained in details in Ref.[38]. Most of the recent debates are linked to the formation of prismatic/basal interfaces. However, to the point view explained in the preamble, the interface features are a consequence of the mechanism and not the cause; and we estimate that it is only when the mechanism is understood that the structure of the dislocations at the interface and in the surrounding matrix can be deduced. Neither the disconnection theory nor the reconstructive model can account for the displacive nature for deformation twinning because the concept of homogenous lattice distortion is voluntarily ignored. To our opinion, it is great time to rehabilitate this concept that was at the core of crystallographic metallurgy for more than a century and that seems to have fallen out of favor for these last decades in some scientific communities.

### 1.3. Objectives of the study

The aim of the present paper is to propose an alternative theory to the "shear", "disconnection" and "reconstructive" theories in the case of the extension twins in magnesium and other hcp metals with c/a ratio close to the ideal ratio. This approach shares with the shear theory the concept of homogenous distortion in order to keep the displacive nature of the transformation. It follows our recent developments in the crystallography of martensitic transformation between the fcc, bcc and hcp phases [23] [24]. The model consists in considering that the atoms are hard spheres of constant size and that the lattice distortion must respect the size of the atoms. The same approach is used here to prove that the habit plane of the $\{10\bar{1}2\}$ extension twins is not fully invariant during the distortion, and that the unit volume and distances in the twinning plane are not constant in the transient states. Such a volume change can indeed be expected because it is known from Kepler (17th-century) that the densest packings of hard spheres are only the hcp and fcc ones, which means that the intermediate states between the initial parent hcp and its final hcp should be less dense [39]. The paper responds to the questions on how the volume change occurs and what is its amplitude. It will be shown that the change of reticular distances are small (a few percent) but larger than the elastic strain of magnesium (~ 0.3 %). This implies that extension twinning in magnesium can't be elastically accommodated and that there is no way for the atoms to follow a simple shear path during the twinning distortion, even by taking into account an elastic deviation from the hard sphere assumption. The model will be used to rigorously determine the orientational and distortional twinning variants. In analogy with martensitic transformations and ferroelectric domains, orientation continuities will be expected between the low-misorientated twin variants. They will be experimentally confirmed by Electron Back Scatter Diffraction (EBSD) maps acquired on a magnesium single crystal. As extension twinning differs from simple shearing, an energy criterion generalizing the Schmid factor will be proposed. It will be shown to predict twinning for "anomalous" conditions, in agreement with some experimental results reported in literature.

## 2. Notations, calculation rules, and experimental details

The three-index notation in the hexagonal system is preferentially chosen for the calculations. The planes will be sometimes written in four-index notation, but mainly to refer to literature. A reader not familiar with the conversion between the three-index and four-index notations can refer to classical textbooks [40]; special attention should be given to the conversion of directions which is more complicated than for planes.

The vectors are noted by bold lowercase letters and the matrices by bold capital letters. In order to calculate the continuous paths of the atoms during the distortion, we made the approximation that the atoms are hard spheres of constant size. The ratio of lattice parameters is then the ideal hcp ratio:

$$\gamma = c/a = \sqrt{\frac{8}{3}} \tag{1}$$

This approximation is a good start for a model of magnesium because this metal is hcp with a ratio $\gamma = 1.623$ very close to the ideal ratio of hard-sphere packing.

We call $\mathbf{B}_{hex} = (\mathbf{a}, \mathbf{b}, \mathbf{c})$ the usual hexagonal basis, and $\mathbf{B}_{ortho} = (\mathbf{x}, \mathbf{y}, \mathbf{z})$ the orthonormal basis represented in Fig. 1 and linked to $\mathbf{B}_{hex}$ by the coordinate transformation matrix $\mathbf{H}_{hex}$:

$$\mathbf{H}_{hex} = [\mathbf{B}_{ortho} \rightarrow \mathbf{B}_{hex}] = \begin{pmatrix} 1 & -1/2 & 0 \\ 0 & \sqrt{3}/2 & 0 \\ 0 & 0 & \gamma \end{pmatrix} \tag{2}$$

In order to follow the displacements of the atoms during extension twinning, some labels are given to the atomic positions, as illustrated in Fig. 1. We note O, the "zero" position that will be let invariant by the distortion. We call X, Y and Z the atomic positions defined by the vectors **OX** = *a* = [100]$_{hex}$, **OY** = *a* + 2*b* = [120]$_{hex}$ and **OZ** = *c* = [001]$_{hex}$. It can be checked with the matrix **H**$_{hex}$ that **OX** = [100]$_{ortho}$, **OY** = [0 $\sqrt{3}$ 0]$_{ortho}$ and **OZ** = [0 0 $\gamma$]$_{ortho}$. The nodes O, X, Y, Z define a non-primitive cell that will be noted XYZ. Other vectors are noted **OS** = **OX** + **OY** = [220]$_{hex}$, **OT** = **OX** + **OZ** = [101]$_{hex}$, **OU** = **OS** + **OZ** = [221]$_{hex}$, **OV** = **OY** + **OZ** = [121]$_{hex}$. The atom at the center of the face (O, X, Y, S) is noted M, and the atom close to the face (O, X, Z, T) is noted N; their position vectors are **OM** = [110]$_{hex}$ and **ON** = [2/3, 1/3, 1/2]$_{hex}$. At the same z-level as N, there are the atoms P and Q given by **OP** = [5/3, 4/3, 1/2]$_{hex}$ and **OQ** = [2/3, 4/3, 1/2]$_{hex}$. Additional positions without atom will be used in the calculations; they are I middle of OX, J middle of OY, and K middle of OZ (Fig. 1d). The bases in which the vectors and matrices are expressed are specified in the text or as indices in the equations.

To describe the crystallography of extension twinning *p*→*t* from a parent crystal *p* to its twin *t*, three important matrices will be used: the distortion matrix, the coordinate transformation matrix, and the correspondence matrix. Let us briefly explain them.

The distortion matrix **D** gives the image **x'** of a vector **x** by a linear distortion: **x'** = **D.x**. The displacement field is given by **x'**-**x** = (**D-I**).**x** where **I** is the 3x3 identity matrix. The deformation matrix is given by its gradient; it is simply the matrix **D-I**. The letter **F** is often used in place of **D** in the textbooks on finite strain theory. The letter **U** is sometimes preferred to **D** to specify that the distortion matrix is symmetric. The distortion matrix can be calculated as follows. The vectors of the initial parent basis are transformed by the distortion into new vectors: $\mathbf{a}_p \rightarrow \mathbf{a'}_p$, $\mathbf{b}_p \rightarrow \mathbf{b'}_p$ and $\mathbf{c}_p \rightarrow \mathbf{c'}_p$. The distortion matrix $\mathbf{D}_{hex}^{p \rightarrow t}$ is the matrix formed by the images $\mathbf{a'}_p$, $\mathbf{b'}_p$ and $\mathbf{c'}_p$ expressed in the initial hexagonal basis, i.e. $\mathbf{D}_{hex}^{p \rightarrow t} = [\mathbf{B}_{hex}^p \rightarrow \mathbf{B}_{hex}'^p] = \mathbf{B}_{hex}'^p$ with $\mathbf{B}_{hex}^p = (\mathbf{a}_p, \mathbf{b}_p, \mathbf{c}_p)$ and $\mathbf{B}_{hex}'^p = (\mathbf{a'}_p, \mathbf{b'}_p, \mathbf{c'}_p)$. The distortion matrix is simply expressed by writing in column the coordinates of $\mathbf{a'}_p$, $\mathbf{b'}_p$ and $\mathbf{c'}_p$ in the basis $\mathbf{B}_{hex}^p$. The crystallographic studies on displacive phase transformations and mechanical twinning often consist in finding the distortion matrices close to the identity matrix in order to minimize the atomic displacements.

If the distortion is known in the basis $\mathbf{B}_{ortho}$, and noted $\mathbf{D}_{ortho}^{p \rightarrow t}$, a formula of coordinate transformation can be used to express it in the basis $\mathbf{B}_{hex}$; it is:

$$\mathbf{D}_{hex}^{p \rightarrow t} = \mathbf{H}_{hex}^{-1} \mathbf{D}_{ortho}^{p \rightarrow t} \mathbf{H}_{hex} \tag{3}$$

with $\mathbf{H}_{hex}$ given by equation (2). Inversely, if the distortion matrix is found in $\mathbf{B}_{hex}$ and it can be written in $\mathbf{B}_{ortho}$ by the formula:

$$\mathbf{D}_{ortho}^{p \rightarrow t} = \mathbf{H}_{hex} \mathbf{D}_{hex}^{p \rightarrow t} \mathbf{H}_{hex}^{-1} \tag{4}$$

The coordinate transformation matrix $\mathbf{T}^{p \rightarrow t}$ allows the change of the coordinates of a fixed vector between the parent and twin bases. It is given by the vectors forming the basis of the twin $\mathbf{B}_{hex}^t = (\mathbf{a}_t, \mathbf{b}_t, \mathbf{c}_t)$ expressed in the parent hexagonal basis, i.e. $\mathbf{T}^{p \rightarrow t} = [\mathbf{B}_{hex}^p \rightarrow \mathbf{B}_{hex}^t]$. This matrix can be calculated from the orientation relationship between the parent and its twin experimentally obtained from TEM or EBSD. The coordinate transformation matrix for the reverse twinning operation is simply $\mathbf{T}^{t \to p} = (\mathbf{T}^{p \to t})^{-1}$.

The correspondence matrix $\mathbf{C}^{t \to p}$ gives the images of the parent basis vectors by the distortion, i.e. $\mathbf{a'}_p$, $\mathbf{b'}_p$ and $\mathbf{c'}_p$, expressed in the twin basis. These images are obtained from the coordinate transformation matrix and the distortion matrix: $(\mathbf{a'}_p, \mathbf{b'}_p, \mathbf{c'}_p)_{/\mathbf{B}^t_{hex}} = \mathbf{T}^{t \to p} (\mathbf{a'}_p, \mathbf{b'}_p, \mathbf{c'}_p)_{/\mathbf{B}^p_{hex}} = \mathbf{T}^{t \to p} \mathbf{B}'^p_{hex} = \mathbf{T}^{t \to p} \mathbf{D}^{p \to t}$. The correspondence matrix is thus:

$$\mathbf{C}^{t \to p} = \mathbf{T}^{t \to p} \mathbf{D}^{p \to t} \tag{5}$$

The correspondence matrix are used to calculate in the twin basis the image by the distortion of a vector of the parent basis, i.e.

$$\mathbf{x'}_{/\mathbf{B}^p_{hex}} = \mathbf{D}^{p \to t} \mathbf{x}_{/\mathbf{B}^p_{hex}} \quad \to \quad \mathbf{x'}_{/\mathbf{B}^t_{hex}} = \mathbf{C}^{p \to t} \mathbf{x}_{/\mathbf{B}^p_{hex}} \tag{6}$$

Most of the symbolic and numerical calculations have been performed with Mathematica. The .nb programs (one for the lattice distortion and another one for the energy criterion) are available in *Supplementary Material 1a* and *1b*, respectively.

Although the paper is mainly theoretical, we used an EBSD map acquired for another study related to compression twinning because this map confirms the intrinsic link between the (86°, **a**) twins and the (90°, **a**) twin modes. A single crystal of magnesium was compressed by 5% at 100°C in direction along the **c**-axis; the experiment was performed on a Gleeble 3800 system. The deformed crystal was mechanically polished with abrasive papers and clothes with diamond particles down to 1 µm, and then electropolished at 12V with an electrolyte made of 85% ethanol with 5% $HNO_3$ and 10% HCl and just taken out of the fridge (10°C). The EBSD map was acquired on a field emission gun (FEG) XLF30 scanning electron microscope (FEI) equipped with the system Aztec (Oxford Instruments). The sample will be shown to exhibit large extension twins. This is surprising because the single crystal was deformed in compression, but this phenomenon was already observed on Mg polycrystalline alloys [41] and Mg single crystals [42] and attributed to the unloading stage.

## 3. Distortion matrices of (90°, a) twinning

### 3.1. Matrix of complete distortion

The case treated in this section and in the next one aims at calculating the stretch component of the lattice distortion. The tilt needed to let $\{10\bar{1}2\}$ plane untilted will be introduced only in a second stage in section 4. This approach is similar to that used for fcc-bcc martensite transformations in steels, where the Bain (stretch) matrix is first calculated, and then an additional rotation is added to compensate the tilt and let a line invariant [23][43]. This section directly calculates the matrix of lattice distortion in the case of a complete transformation, without considering the continuous path that leads to it. The calculation of the continuous expression of the stretch matrix will be the subject of the section 3.2.

In order to obtain the stretch component of the extension twinning, we consider the case in which the axis ***a*** remains invariant, the axis ***a+2b*** of the parent crystal is transformed into the axis ***c*** of the twin, and the axis ***c*** of the parent is transformed into the axis ***a+2b*** of the twin:

$$\mathbf{a}'_p = \mathbf{a}_t, \quad \mathbf{a}'_p + 2\mathbf{b}'_p = \mathbf{c}_t, \quad \mathbf{c}'_p = \mathbf{a}_t + 2\mathbf{b}_t \tag{7}$$

which by linear combination gives

$$\mathbf{a}'_p = \mathbf{a}_t, \quad \mathbf{b}'_p = \tfrac{1}{2}(\mathbf{c}_t - \mathbf{a}_t), \quad \mathbf{c}'_p = \mathbf{a}_t + 2\mathbf{b}_t \tag{8}$$

Thus, the correspondence matrix $\mathbf{C}^{t \to p}$ of extension twinning expressed in the basis $\mathbf{B}_{hex}$ is:

$$\mathbf{C}^{t \to p}_{hex} = \begin{pmatrix} 1 & -1/2 & 1 \\ 0 & 0 & 2 \\ 0 & 1/2 & 0 \end{pmatrix} \tag{9}$$

The coordinate transformation matrix $\mathbf{T}^{p \to t}_{hex}$ is given by the vectors of the twin basis expressed in the parent basis. In the present case, this matrix is similar to the correspondence matrix, excepted that now the metrics appears:

$$\mathbf{T}^{p \to t}_{hex} = \begin{pmatrix} 1 & -1/2 & \frac{\gamma}{\sqrt{3}} \\ 0 & 0 & \frac{2\gamma}{\sqrt{3}} \\ 0 & \frac{\sqrt{3}}{2\gamma} & 0 \end{pmatrix} = \begin{pmatrix} 1 & -1/2 & \frac{2\sqrt{2}}{3} \\ 0 & 0 & \frac{4\sqrt{2}}{3} \\ 0 & \frac{3}{4\sqrt{2}} & 0 \end{pmatrix} \tag{10}$$

$$\Rightarrow \mathbf{T}^{p \to t}_{hex} = \begin{pmatrix} 1 & -1/2 & \frac{2\sqrt{2}}{3} \\ 0 & 0 & \frac{4\sqrt{2}}{3} \\ 0 & \frac{3}{4\sqrt{2}} & 0 \end{pmatrix} \quad \text{for } \gamma = \sqrt{\frac{8}{3}} \tag{11}$$

The coordinate transformation matrix transforms the hexagonal parent basis into the hexagonal twin basis. The equivalent matrix that transforms the orthonormal basis of the parent crystal into the orthonormal basis of the twin crystal is given by

$$\overline{\mathbf{R}}^{p \to t}_{ortho} = \mathbf{H}_{hex} \mathbf{T}^{p \to t}_{hex} \mathbf{H}^{-1}_{hex} = \begin{pmatrix} 1 & 0 & 0 \\ 0 & 0 & 1 \\ 0 & 1 & 0 \end{pmatrix} \tag{12}$$

which, as expected, is a mirror symmetry across the plane Y+Z=0 or equivalently an improper rotation of axis **OX** = $\mathbf{a}_p = \mathbf{a}_t$ and angle 90°.

The distortion matrix $\mathbf{U}^{p \to t}_{hex}$ is obtained from the correspondence and coordinate transformation matrices by using the relation (5):

$$U_{hex}^{p \to t} = T_{hex}^{p \to t} C_{hex}^{t \to p} = \begin{pmatrix} 1 & -\frac{1}{2} + \frac{\gamma}{2\sqrt{3}} & 0 \\ 0 & \frac{\gamma}{\sqrt{3}} & 0 \\ 0 & 0 & \frac{\sqrt{3}}{\gamma} \end{pmatrix} \quad (13)$$

$$\Rightarrow U_{hex}^{p \to t} = \begin{pmatrix} 1 & -\frac{1}{2} + \frac{\sqrt{2}}{3} & 0 \\ 0 & \frac{2\sqrt{2}}{3} & 0 \\ 0 & 0 & \frac{3}{2\sqrt{2}} \end{pmatrix} \quad for\ \gamma = \sqrt{\frac{8}{3}} \quad (14)$$

This matrix is given in the basis $\mathbf{B}_{hex}^p$. It can be written in the basis $\mathbf{B}_{ortho}^p$ by using equation (4):

$$U_{ortho}^{p \to t} = H_{hex} U_{hex}^{p \to t} H_{hex}^{-1} = \begin{pmatrix} 1 & 0 & 0 \\ 0 & \frac{\gamma}{\sqrt{3}} & 0 \\ 0 & 0 & \frac{\sqrt{3}}{\gamma} \end{pmatrix} \quad (15)$$

$$\Rightarrow U_{ortho}^{p \to t} = \begin{pmatrix} 1 & 0 & 0 \\ 0 & \frac{2\sqrt{2}}{3} & 0 \\ 0 & 0 & \frac{3}{2\sqrt{2}} \end{pmatrix} \approx \begin{pmatrix} 1 & 0 & 0 \\ 0 & 0.94 & 0 \\ 0 & 0 & 1.06 \end{pmatrix} \quad for\ \gamma = \sqrt{\frac{8}{3}} \quad (16)$$

The general formula (15) proves that the twin is an extension twin (along the **z**-axis) for $\gamma \leq \sqrt{3}$, and a contraction twin for $\gamma \geq \sqrt{3}$, as already noticed by Yoo [44].

Its numerical values (16) obtained for a perfect hcp packing were already reported in literature [35][37]. This matrix is diagonal; it is a stretch equivalent to the Bain distortion known in martensitic transformations. Its values are x'/x, y'/y, z'/z; they mean that the distortion lets the **x**-axis invariant, shorten the **y**-axis by Δy/y = -6% and extends the **z**-axis by Δz/z = +6%. The ratio of contraction y'/y is exactly the inverse of the ratio of extension z'/z. One can now raise the question: how the two ratios evolve *during* the lattice distortion? Do they follow this inverse relation continuously, which would mean that the volume is unchanged during the lattice distortion? To answer this question one needs to go deeper into the details of the distortion by considering the atomic displacements.

### 3.2. Matric of continuous distortion

In the rest of the paper, it will be assumed that the atoms are hard spheres and that these atoms are perfectly packed. The c/a ratio is thus fixed at $\gamma = \sqrt{\frac{8}{3}}$.

The mechanistic reason for the simultaneous y'/y contraction and z'/z extension is the coordinated displacement of the M and N atoms. Indeed, these atoms, taken as hard spheres, are in "contact" and keep contact during their movements. The atom M initially in the basal plane (O, X, Y, S) = $(001)_p$ goes out of the plane such that after twinning this plane is transformed into the prismatic plane (O, X, Y, S) = (010)$_t$, and the atom N goes toward the prismatic plane (010)$_p$ such that after twinning this plane is transformed into the basal plane (001)$_t$, as illustrated in Fig. 1 and in Fig. 2a and b. During the displacements of the atoms M and N, the distance OZ increases and the distance OY decreases. It is assumed that the atoms O and X do not move, and that the atoms M and N keep contact with the atoms O and X, and between each other during their displacements. The trajectories of the atoms M and N can then be described by a unique parameter, which is the angle $\eta$ made by the vector **IM** with the basal plane (O, X, Y, S) (Fig. 1 and Fig. 2). Let us call $\Phi$ the angle between the direction **IN** with the basal plane (O, X, Y, S), i.e. $\Phi = ArcCos\left(\frac{1}{3}\right) \approx 70.5°$. During extension twinning, the angle $\eta$ increases from the start value $\eta_s = 0$ to the finish value $\eta_f = \frac{\pi}{2} - \Phi \approx 19.47°$. The coordinates of the atoms M and N in the orthonormal basis $\mathbf{B}_{ortho}$ are

$$\mathbf{OM} = \begin{pmatrix} 1/2 \\ \frac{\sqrt{3}}{2} Cos(\eta) \\ \frac{\sqrt{3}}{2} Sin(\eta) \end{pmatrix} \text{ and } \mathbf{ON} = \begin{pmatrix} 1/2 \\ \frac{\sqrt{3}}{2} Cos(\eta + \Phi) \\ \frac{\sqrt{3}}{2} Sin(\eta + \Phi) \end{pmatrix} \quad (17)$$

In the present case, it is supposed that **OY'** remains parallel to **OY** during the transformation. The atom M moves such that it keeps contact with the atoms O, X and Y. Thus, as shown in Fig. 2d, the point J in the middle of OY keeps the same y-coordinate as M, i.e. OY = 2 OJ = $\sqrt{3} Cos(\eta)$. It is also assumed that OZ' remains parallel to OZ. The atom N moves such that it keeps contact with the atoms O, X and Z. Thus, as shown in Fig. 2d, the point K in the middle of OZ keeps the same z-coordinate as N, i.e. OZ = 2 OK = $\sqrt{3} Sin(\eta + \Phi)$. Let us write the vectors **OX**, **OY** and **OZ** forming the basis $\mathbf{B}_{XYZ}$ of the XYZ cell in the orthonormal basis $\mathbf{B}_{ortho}$. The coordinate transformation matrix between these bases is

$$[\mathbf{B}_{ortho} \rightarrow \mathbf{B}_{XYZ}(\eta)] = \mathbf{B}_{XYZ}(\eta) = \begin{pmatrix} 1 & 0 & 0 \\ 0 & \sqrt{3} Cos(\eta) & 0 \\ 0 & 0 & \sqrt{3} Sin(\eta + \Phi) \end{pmatrix} \quad (18)$$

The continuous distortion matrix at each step $\eta$ of the distortion is given in the basis $\mathbf{B}_{ortho}$ by the matrix $\mathbf{U}_{ortho}^{p \rightarrow t}(\eta) = \mathbf{B}_{XYZ}(\eta) \cdot \mathbf{B}_{XYZ}^{-1}(\eta_s)$ (see equation 1 of Ref.[1]). The calculation leads to

$$\mathbf{U}_{ortho}^{p \rightarrow t}(\eta) = \begin{pmatrix} 1 & 0 & 0 \\ 0 & Cos(\eta) & 0 \\ 0 & 0 & \frac{3}{2\sqrt{2}} Sin(\eta + \Phi) \end{pmatrix} \quad (19)$$

In order to get simpler expressions, we introduce the variable $\kappa = Sin(\eta)$. The start value becomes $\kappa_s = 0$ and the finish value $\kappa_f = 1/3$. The distortion matrix is now a function of $\kappa$:

$$\mathbf{U}_{ortho}^{p \rightarrow t}(\kappa) = \begin{pmatrix} 1 & 0 & 0 \\ 0 & \sqrt{1-\kappa^2} & 0 \\ 0 & 0 & \frac{\kappa}{2\sqrt{2}} + \sqrt{1-\kappa^2} \end{pmatrix} \quad (20)$$

The matrix of complete transformation is given for $\kappa_f = 1/3$; it can be checked that it is the matrix already given in equation (16). This proves that the ratio of contraction y'/y is the inverse of the ratio of extension z'/z *only* in the final state, but not during the intermediate states of the distortion.

The distortion matrix can be written in the hexagonal basis $\mathbf{B}_{hex}$ by using equation (3):

$$\mathbf{U}_{hex}^{p\to t}(\kappa) = \begin{pmatrix} 1 & \dfrac{-1+\sqrt{1-\kappa^2}}{2} & 0 \\ 0 & \sqrt{1-\kappa^2} & 0 \\ 0 & 0 & \dfrac{\kappa}{2\sqrt{2}} + \sqrt{1-\kappa^2} \end{pmatrix} \quad (21)$$

The ratio of volume change $\mathcal{V}'/\mathcal{V}$, where $\mathcal{V}$ is the initial volume of the XYZ cell and $\mathcal{V}'$ is the volume of the distorted cell, is directly given by the determinant of the distortion matrix:

$$\frac{\mathcal{V}'}{\mathcal{V}}(\kappa) = det(\mathbf{U}_{hex}^{p\to t}) = det(\mathbf{U}_{ortho}^{p\to t}) = 1 - \kappa^2 + \frac{\sqrt{2}}{4}\kappa\sqrt{1-\kappa^2} \quad (22)$$

The curve $\mathcal{V}'/\mathcal{V}$ is presented in Fig. 3. The maximum of volume change, close to 1.0303 is obtained for the intermediate value $\kappa_i \approx 0.1691$ ($\eta_i \approx 9.73°$). It proves that, although it returns to its initial value when the distortion is complete $\frac{\mathcal{V}'}{\mathcal{V}}(\kappa_f) = 1$, the volume is not constant during the twinning process. Thus, it is proved that extension twinning cannot be obtained at constant volume with a hard-sphere model. A similar conclusion was drawn on fcc-fcc mechanical twinning in a previous study (sections 3 and 7.5 of Ref.[1]).

The matrix $\mathbf{U}_{hex}^{p\to t}(\kappa)$ gives the stretch (Bain) matrix during the process of extension twinning. The atoms at the first corners of the XYZ cell (i.e. O, X, Y, Z) follow a trajectory whose equation in the basis $\mathbf{B}_{ortho}$ is directly given by this matrix, with $\kappa$ continuously varying from $\kappa_s = 0$ to $\kappa_f = 1/3$. The atoms M, N, Q inside the XYZ cell do not follow the same trajectory; they "shuffle". The trajectory equation of M and N is deduced from equation (17). Actually, the initial position vectors in the basis $\mathbf{B}_{ortho}$, which are $\mathbf{OM} = \left[\frac{1}{2}, \frac{\sqrt{3}}{2}, 0\right]$, $\mathbf{ON} = \left[\frac{1}{2}, \frac{\sqrt{3}}{6}, \frac{\sqrt{6}}{3}\right]$, $\mathbf{OQ} = \left[0, \frac{2\sqrt{3}}{3}, \frac{\sqrt{6}}{3}\right]$, are all rotated by the same "shuffle rotation" that is simply

$$\mathbf{R}_{shuffle}(\eta) = \begin{pmatrix} 1 & 0 & 0 \\ 0 & Cos(\eta) & -Sin(\eta) \\ 0 & Sin(\eta) & Cos(\eta) \end{pmatrix} \quad (23)$$

Consequently, in the XYZ cell, the corner atoms O, X, Y, Z, S, T, U, V (each of them count for 1/8 in the cell) follow the distortion (20), and the atom M (counts for ½), N (counts for 1), Q (counts for ½) follow the shuffling rotation (23). This means that twinning is obtained with 1/3 of distortion and 2/3 of shuffle.

Some movies displaying the atomic displacements in a magnesium crystal during the twinning transformation were computed with VPython. The first movie (*Supplementary Material 2*) shows the transformation of a Bravais unit cell, and the second movie (*Supplementary Material 3*) shows the transformation of a 4x4x4 XYZ supercell. Three snapshots, taken from these two movies at initial, intermediate and final states, are extracted and given in Fig. 4 and Fig. 5.

## 4. Distortion matrices of (86°, a) twinning

### 4.1. Compensating angle and matrix of continuous distortion

The distortion matrix (20) is diagonal and none of its values equals 1, which means that there is no direction or plane that is let invariant. The orientation relationship associated with the distortion matrix by equation (12) is a (90°, *a*) (improper) rotation, as it was observed in submicron-sized Mg single-crystal [35]. In bulk magnesium, it is known that extension twins let "invariant" the $\{10\bar{1}2\}$ planes, which implies that the orientation relationship is (86°, *a*) and not (90°, *a*). In order to make this plane untilted during the lattice distortion, a rotation should be added to the matrix (20), in the same way as a rotation is added to the Bain tensor in fcc-bcc transformations in steels (see for example ref. [43]). Since the direction **OX** = *a* is already invariant, the $(0\bar{1}12)$ plane can be continuously maintained untilted by compensating the rotation of the direction **OV** = $[121]_{hex}$ ∥ $[0\ 3\ 2\sqrt{2}]_{ortho}$. Let us call $\xi$ the angle (**OV**, **OV'**) illustrated in Fig. 6a. The cosine of this angle $C_\xi$ is calculated by the scalar product $\mathbf{OV}_{ortho} \cdot \mathbf{U}^{p \to t}_{ortho} \mathbf{OV}_{ortho}$ using the matrix (20). It is

$$C_\xi(\kappa) = \frac{2\sqrt{2}\kappa + 17\sqrt{1-\kappa^2}}{\sqrt{17}\sqrt{17 - 16\kappa^2 + 4\kappa\sqrt{2 - 2\kappa^2}}} \tag{24}$$

The angle $\xi$ varies from 0 to 3.4° during the diagonal distortion. The rotation matrix that compensates the angle $\xi$ is thus given in the basis $\mathbf{B}_{ortho}$ by

$$\mathbf{R}(\kappa) = \begin{pmatrix} 1 & 0 & 0 \\ 0 & C_\xi(\kappa) & \sqrt{1 - C_\xi(\kappa)^2} \\ 0 & -\sqrt{1 - C_\xi(\kappa)^2} & C_\xi(\kappa) \end{pmatrix} \tag{25}$$

We point out here that only the rotation of the direction **OV** can be cancelled, but not its length change. Indeed, the distance $\|\mathbf{OV}(\kappa)\| = \|\mathbf{U}^{p \to t}_{ortho}(\kappa)\mathbf{OV}\|$ calculated from the matrix (20) is not constant. The ratio of the length OV' divided by its initial value OV is

$$\frac{OV'}{OV}(\kappa) = \frac{\sqrt{17 - 16\kappa^2 + 4\kappa\sqrt{2 - 2\kappa^2}}}{\sqrt{17}} \tag{26}$$

Its graph is given in Fig. 6b. This proves that, although the length OV returns to its initial value when the distortion is complete, the distance is not constant during the twinning process. Consequently, although the plane $(0\bar{1}12)$ remains untilted and is eventually restored, this plane is *not* fully invariant during the process. Strictly speaking, one should say that the twinning plane $(0\bar{1}12)$ is *globally* invariant (untilted and restored).

The distortion matrix that lets the $(0\bar{1}12)$ plane untilted is $\mathbf{D}^{p \to t}_{ortho}(\kappa) = \mathbf{R}(\kappa)\,\mathbf{U}^{p \to t}_{ortho}(\kappa)$. The calculations show that

$$\mathbf{D}_{ortho}^{p \to t}(\kappa) = \begin{pmatrix} 1 & 0 & 0 \\ 0 & \dfrac{17 - 17\kappa^2 + 2\kappa\sqrt{2 - 2\kappa^2}}{\sqrt{17}\, d(\kappa)} & \dfrac{3\kappa(\kappa + 2\sqrt{2 - 2\kappa^2})}{2\sqrt{34}\, d(\kappa)} \\ 0 & -\dfrac{3}{\sqrt{17}} \dfrac{\kappa\sqrt{1 - \kappa^2}}{d(\kappa)} & \dfrac{68 - 64\kappa^2 + 25\kappa\sqrt{2 - 2\kappa^2}}{4\sqrt{17}\, d(\kappa)} \end{pmatrix} \quad (27)$$

with $d(\kappa) = \sqrt{17 - 16\kappa^2 + 4\kappa\sqrt{2 - 2\kappa^2}}$

The matrix $\mathbf{D}_{ortho}^{p \to t}(\kappa)$ is the full form of the lattice distortion during extension twinning. The displacements of the atoms at the corners of the XYZ cell (i.e. O, X, Y, Z) follow a trajectory whose equation in the basis $\mathbf{B}_{ortho}$ is directly given by this matrix, with κ continuously varying from $\kappa_s = 0$ to $\kappa_f = 1/3$. The atoms M, N, Q shuffle in the XYZ cell. Their trajectories are given by the "shuffling rotation" (23) compensated of the angle ξ, i.e.

$$\mathbf{R}_{shuffle}(\eta) = \begin{pmatrix} 1 & 0 & 0 \\ 0 & Cos(\eta - \xi) & -Sin(\eta - \xi) \\ 0 & Sin(\eta - \xi) & Cos(\eta - \xi) \end{pmatrix} \quad (28)$$

with ξ function of η by equation (24).

As the distortion matrix (27) is the product of a rotation matrix by the stretch matrix (20), the volume change is the same as the one given in equation (22). The changes of the unit volume and of the OV distance in $(0\bar{1}12)$ plane mean that simple shear matrices are not the most appropriate tool to describe the continuous paths of the atoms.

A movie of the transformation of 4x4x4 XYZ cell is reported in *Supplementary Material 4*. Three snapshots taken at initial, intermediate and final states are extracted and shown in Fig. 7. It can be checked that the $(0\bar{1}12)$ plane is not tilted during the transformation. Its intra-planar distortion in the transient states due to the change of length OV (+1.4%) is hardly perceptible.

### 4.2. Complete form of the distortion and orientation matrices

When the transformation is complete, the distortion matrix takes the value

$$\mathbf{D}_{ortho}^{p \to t} = \mathbf{D}_{ortho}^{p \to t}(1/3) = \begin{pmatrix} 1 & 0 & 0 \\ 0 & \dfrac{16}{17} & \dfrac{3}{34\sqrt{2}} \\ 0 & -\dfrac{2\sqrt{2}}{51} & \dfrac{18}{17} \end{pmatrix} \quad (29)$$

In the hexagonal basis $\mathbf{B}_{hex}$, it becomes, by using equation (3),

$$\mathbf{D}_{hex}^{p \to t} = \begin{pmatrix} 1 & -\dfrac{1}{34} & \dfrac{1}{17} \\ 0 & \dfrac{16}{17} & \dfrac{2}{17} \\ 0 & -\dfrac{1}{34} & \dfrac{18}{17} \end{pmatrix} \quad (30)$$

In the reciprocal space, one must take the inverse of the transpose:

$$(\mathbf{D}_{ortho}^{p \to t})^* = (\mathbf{D}_{ortho}^{p \to t})^{-T} = \begin{pmatrix} 1 & 0 & 0 \\ 0 & \dfrac{18}{17} & \dfrac{2\sqrt{2}}{51} \\ 0 & -\dfrac{3}{34\sqrt{2}} & \dfrac{16}{17} \end{pmatrix} \quad (31)$$

$$(\mathbf{D}_{hex}^{p \to t})^* = (\mathbf{D}_{hex}^{p \to t})^{-T} = \begin{pmatrix} 1 & 0 & 0 \\ \dfrac{1}{34} & \dfrac{18}{17} & \dfrac{1}{34} \\ -\dfrac{1}{17} & -\dfrac{2}{17} & \dfrac{16}{17} \end{pmatrix} \quad (32)$$

The distortion matrix in the direct space and hexagonal basis given by equation (30) has only one eigenvalue equal to 1 and an infinity of eigenvectors that are all linear combinations of the two vectors **OX** = [100]$_{hex}$ and **OV** = [121]$_{hex}$. This means that the distortion matrix (30) is a simple shear matrix on the plane $(0\bar{1}2)_{hex}$. The shear coefficient $s$ is the tangent of the angle made by the vector **n** normal of the shear plane with its image. It is easily calculated with $\mathbf{n} = [0, -\gamma, \sqrt{3}]_{ortho}$ and its image by the matrix (29); it is $s = \dfrac{1}{6\sqrt{2}} \approx 0.118$, as expected from the theoretical value of shear [9]. The shear vector **s** is

$$\mathbf{s} = (\mathbf{D}_{ortho}^{p \to t} - \mathbf{I}) \cdot \dfrac{\mathbf{n}}{\|\mathbf{n}\|} = \dfrac{1}{\sqrt{34}} \left[0, \dfrac{1}{2}, \dfrac{\sqrt{2}}{3}\right]_{ortho} = \dfrac{1}{2\sqrt{102}} [121]_{hex} \quad (33)$$

The vector **s** makes an angle of 43.31° with the basal plane. It can be checked that the vector **s** belongs to the "shear" plane $(0\bar{1}2)_{hex}$. The distortion matrix expressed in the reciprocal space is given by equation (32); it has only one eigenvalue equal to 1 and an infinity of eigenvectors that are all linear combinations of the two vectors $(10\bar{1})_{hex}$ and $(\bar{2}10)_{hex}$. This means that, in addition to the shear plane $(0\bar{1}2)_{hex}$, all the planes that contain the shear vector **s** are also invariant, as expected from a shear matrix. These calculations show that the continuous approach allows us to find the classical result (a shear matrix) in the special case where only the initial and final states are considered. To our knowledge, it is the first time that a continuous analytical form of the distortion matrix is given to model the twinning transformation process.

The orientation of the twin is equal to that of the (90°, **a**) twins corrected of the 3.4° angle, i.e. it is (86.6°, **a**), simply noted in the paper (86°, **a**). This result can also be obtained by calculating the coordinate transformation matrix

$$\mathbf{T}_{hex}^{p \to t} = \mathbf{D}_{hex}^{p \to t} \cdot (\mathbf{C}_{hex}^{t \to p})^{-1} \quad (34)$$

with $\mathbf{D}_{hex}^{p \to t}$ given by the equation (30) and $\mathbf{C}_{hex}^{t \to p}$ by the equation (9). In order to obtain the rotation matrix, the matrix $\mathbf{T}_{hex}^{p \to t}$ must be composed with the mirror symmetry **M** across the basal plane and then be expressed in the orthonormal basis. The result is

$$\mathbf{R}^{p\to t} = \mathbf{H}_{hex}\mathbf{M}_{hex}\,\mathbf{T}^{p\to t}_{hex}\,\mathbf{H}^{-1}_{hex} = \begin{pmatrix} 1 & 0 & 0 \\ 0 & \dfrac{1}{17} & \dfrac{12\sqrt{2}}{17} \\ 0 & -\dfrac{12\sqrt{2}}{17} & \dfrac{1}{17} \end{pmatrix} \quad (35)$$

The matrix $\mathbf{R}^{p\to t}$ is indeed a rotation around the **a**-axis of angle ArcCos(1/17) = 86.6°.

## 5. Orientational and distortional variants

The orientational variants are the twin variants that are similarly orientated with the parent crystal. At maximum, there are as many variants as symmetries in the point group of the parent phase $\mathbb{G}$, but often the number is lower due to the fact that, for some special orientation relationships $\mathbf{T}^{p\to t}$, some symmetries are common to the parent crystal and its twins. These common symmetries form a subgroup of $\mathbb{G}$, called "intersection group" $\mathbb{H}$. In the case of deformation twinning

$$\mathbb{H} = \mathbb{G} \cap \mathbf{T}^{p\to t}\,\mathbb{G}\,(\mathbf{T}^{p\to t})^{-1} \quad (36)$$

The orientational twin variants $t_i$ can mathematically identified to the left-cosets based on $\mathbb{H}$, i.e. $g_i\mathbb{H}$, and their number is thus given by the Lagrange's formula; it is the order of the parent point group $\mathbb{G}$ divided by the order of the intersection group $\mathbb{H}$. The distinct orientations are

$$\mathbf{T}^{p\to t_i} = g_i\mathbf{T}^{p\to t} \text{ with } g_i \text{ a symmetry arbitrarily chosen in the coset } g_i\mathbb{H} \quad (37)$$

The misorientations between the variants, called "operators", are identified the double-cosets based on the intersection group, $\mathbb{H}g_i\mathbb{H}$, and their number is given by the Burnside's formula. The variants and their operators form an algebraic structure called groupoid. More details on these concepts are given in Ref. [45].

The distortional variants are defined similarly as the orientational variants, but by replacing the orientation relationship matrix $\mathbf{T}^{p\to t}$ by the distortion matrix $\mathbf{D}^{p\to t}$ [24]. The intersection group is

$$\mathbb{K} = \mathbb{G} \cap \mathbf{D}^{p\to t}\,\mathbb{G}\,(\mathbf{D}^{p\to t})^{-1} \quad (38)$$

The distortional variants are the left-cosets based on the subgroup $\mathbb{K}$, i.e. $g_i\mathbb{K}$. The matrices of the distinct distortions are

$$\mathbf{D}^{p\to t_i} = g_i\,\mathbf{D}_0^{\gamma\to\alpha}(g_i)^{-1} \text{ with } g_i \text{ a symmetry arbitrarily chosen in the coset } g_i\mathbb{K} \quad (39)$$

The number of distortional variants is also given by the Lagrange's formula. It is not always equal to the number of orientational variants (see the case of fcc-hcp transition in Ref. [24]).

The point group of hcp metals is $\mathbb{G} = 6/mmm$; it is constituted of 24 symmetry matrices, and these 3x3 matrices, when expressed in the hexagonal frame, are made of 0, 1 or -1. The calculations of the extension twinning variants are direct. They were done geometrically and algebraically. The results were also confirmed by the software GenOVa [46] (*Supplementary Material 5*).

In the case of the (90°, **a**) twin, the intersection group $\mathbb{H}$ is constituted by the identity, the inversion, the 2-fold rotation around the **a**-axis, the mirror symmetry across the basal plane, and their combinations. This geometrical result is also numerically obtained by using formula (36) with the coordinate transformation matrix $\mathbf{T}^{p \to t}$ given in equation (11). The number of twin variants is thus 24/8 = 3. For distortional variants, the intersection group $\mathbb{K}$ is calculated by using equation (38) with the distortion matrix (14). The calculations prove that $\mathbb{K} \equiv \mathbb{H}$, and consequently, there are also three distortional variants.

In the case of the (86°, **a**) extension twins, the mirror symmetry across the basal plane is not anymore a common symmetry and consequently the intersection group $\mathbb{H}$ is reduced to four symmetry matrices: the identity, the inversion, the 2-fold rotation around the **a**-axis and the mirror symmetry across the plane normal to the **a**-axis, i.e. the $(2\bar{1}\bar{1}0)$ plane. This result is also obtained numerically by using formula (36) with the coordinate transformation matrix $\mathbf{T}^{p \to t}$ given in equation (34). The number of twin variants is thus 24/4 = 6. Here again, the calculations prove that $\mathbb{K} \equiv \mathbb{H}$, and thus, the number of distortional variants is also equal to six. The six twin variants can be grouped into 3 pairs of low-disorientated variants. For example, a pair is constituted by the variants $t_1 = \mathbb{H}$ and $t_5 = \boldsymbol{g}_5 \mathbb{H}$ (see *Supplementary Material 5*). These variants are linked by the operator $O_3$ which contains the rotation of 7.4° around the **a**-axis. The distortion matrices associated with these variants are

$$\mathbf{D}^{p \to t1}_{hex} = \begin{pmatrix} 1 & -\frac{1}{34} & \frac{1}{17} \\ 0 & \frac{16}{17} & \frac{2}{17} \\ 0 & -\frac{1}{34} & \frac{18}{17} \end{pmatrix} \text{ and } \mathbf{D}^{p \to t5}_{hex} = \boldsymbol{g}_5 \mathbf{D}^{p \to t1}_{hex} \boldsymbol{g}_5^{-1} = \begin{pmatrix} 1 & -\frac{1}{34} & -\frac{1}{17} \\ 0 & \frac{16}{17} & -\frac{2}{17} \\ 0 & \frac{1}{34} & \frac{18}{17} \end{pmatrix} \quad (40)$$

The difference between the two matrices is small:

$$\mathbf{D}^{p \to t1}_{hex} - \mathbf{D}^{p \to t5}_{hex} = \begin{pmatrix} 0 & 0 & \frac{2}{17} \\ 0 & 0 & \frac{4}{17} \\ 0 & -\frac{1}{17} & 0 \end{pmatrix} \quad (41)$$

This result can be explained by the fact that the two (86°, **a**) variants $t_1$ and $t_5$ are derived from the same (90°, **a**) variant by the same distortion matrix (14) and differ only by the sign of the correction angle. An analogy with ferroelectrics and martensitic transformations will be discussed in the next section.

# 6. Discussion

## 6.1. A unique mechanism for the 90° and 86° extension twin domains

The classical "shear" model of deformation twinning is based on the lattice correspondences resulting from simple shears; but the atomic displacements are not really taken into account, they are inferred as "shuffles" only after determining the lattice shear. The present model reverses the order of thinking. The atomic trajectories are calculated in order to restore the crystal structure in a special orientation relationship. The magnesium atoms are assumed to be hard spheres of constant size and their displacements are chosen as low as possible. These assumptions constrain the calculations such that only one parameter, a distortion angle $\eta$, is sufficient to follow the trajectories of all the atoms during the process of extension twinning. The distortion matrix is calculated as an analytical function of the angular parameter $\eta$ only once the atomic displacements are known.

First, the calculations were performed by assuming that the final parent/twin orientation is (90°, ***a***), as observed by Liu *et al.* in submicron-sized Mg pillars [35][36]. In such as case, the distortion matrix is diagonal, i.e. it corresponds to a stretch deformation. However, contrarily to the interpretation based on unit cell reconstruction, our model is displacive, as expected for a process implying rapid and collective atomic displacements. In a second step, we showed that the distortion matrix that lets "invariant" the $(0\bar{1}12)$ plane differs from the stretch form only by a small "compensating" rotation of angle that varies continuously from 0 to of 3.4°. In other words, the diagonal distortion matrix related to Liu *et al.*'s observations is the symmetric matrix that can be extracted by polar decomposition from the usual twinning matrix, as already shown by Ostapovest *et al.* [37], in a similar way that the Bain tensor appears as a component of the lattice distortion in the fcc-bcc martensitic transformation in steels [23]. However, we don't believe that the Liu *et al.*'s observations result from an average of two twinning shear on different conjugate $\{10\bar{1}2\}$ planes, as proposed by Ostapovest *et al*. We think that the observed (90°, ***a***) twins and related stretching observed in the magnesium nano-pillars actually result from a real "natural" mechanism that is free to appear due the small size of the sample. The interface strains required for the basal/prism and prism/basal interfaces (-6% along ***a***+2***b*** and + 6% along ***c***) are probably more easily accommodated in the "free" submicron-sized pillar than in the bulk samples. In bulk single crystal or polycrystalline magnesium, such strains are probably too high and need to be reduced by a slight crystal rotation of 3.4° that permits to let the $\{10\bar{1}2\}$ habit plane untilted. The distortion mechanism in the submicron pillars and in bulk samples are actually very similar because the distortion matrices (14) and (30) have close numerical values.

## 6.2. Analogies and expectations

To our point of view, there is a strong analogy between the extension twins in magnesium and the orientational domains in ferroelectrics. Let us explain. If, for sake of simplicity, we represent the cubic→tetragonal transition in a ferroelectric crystal by a square→rectangle distortion along the **x** and **y**-axes, four domains misoriented by 90° and 180° rotations should be formed, as illustrated in Fig. 8a. However, experience shows that the domains are actually oriented such that they share one of their two diagonals. There are thus actually height domains, and among them, the 90° misorientation angle is lost and replaced by 90°-2$\delta$, with $\delta$ = ArcTan($a/c$), and $a$ and $c$ the lattice parameters of the tetragonal phase, as shown in Fig. 8b. The angle $\delta$ compensates the rotation of the diagonal induced by the tetragonal distortion; it is called "obliquity". The concept of obliquity dates from Friedel's work in 1920 [47]. It is a major parameter to define twins in minerals and ferroelectrics [48]-[50]. For example, in barium titanate, the tetragonality is close to 1% and $\delta$ is close to 0.6° [45]. The fact that the diagonal is maintained invariant is explained by strain minimization and compatibility conditions [51][52]. There is a crystallographic equivalence between i) the 86°= 90°-$\xi$ rotation between a $\{0\bar{1}12\}$ twin and its parent in magnesium, and ii) the 90°-2$\delta$ rotation between two ferroelectric domains. Following the vocabulary used for ferroelectrics, we would say that the tetragonal distortion observed by Liu *et al.* in the submicron-sized pillars is a "spontaneous" mechanism, and that the angle $\xi$ required to maintain untilted the $\{10\bar{1}2\}$ planes and form the (86°, **a**) twins in bulk magnesium is an obliquity angle. A clear analogy exist between the two low-misoriented (86°, **a**) variants derived from the same (90°, **a**) variant, and the domains $D_z^+$ and $D_z^-$ derived from the domain $D_z$, as illustrated in Fig. 9a.

Before showing the EBSD results, we would like to explain the aims of our experiments. We came to study the mechanism of fcc-bcc martensitic transformation in order to explain the continuous features observed in the X-ray and EBSD pole figures [53]-[55]. To our opinion, these features are the plastic trace let by the lattice distortion. They can be imaged a) as the consequences of the back-stresses created by the geometrically necessary dislocations (GND) generated in the surrounding matrix by the lattice distortion [54], or b) as the result of the growth of the product phase inside the surrounding parent matrix that has been rotated by the GNDs [55]. For example, continuous features were observed between the two variants in the pairs of low-misoriented Kurdjumov-Sachs (KS) variants, and the other Nishiyama-Wasserman (NW) and Pitsch variants. The 24 KS variants are formed according to a low-symmetry OR (intersection group of order 2) and the 12 NW and the 12 Pitsch variants are formed according to a higher-symmetry OR (intersection group of order 4). As we consider that deformation twinning is a displacive phase transformation, similar features are expected for the extension twins in magnesium. More specifically, it is expected that the back-stresses generated by the twin distortions inside the surrounding matrix will create a continuous rotation between the two (low-symmetry OR) $D_z^-$ and $D_z^+$ variants, re-orienting them back to the initial orientation of the parent lattice and to the (high-symmetry OR) $D_z$ variant. A continuous rotation of angle 2$\xi$ around the **a**-axis is expected between the $D_z^-$ and $D_z^+$ variants, as schematically shown in Fig. 9b.

### 6.3. Experimental confirmation of the link between the (86°, a) and the (90°, a) twins.

The possibility to form concomitantly pairs of low-misoriented variants and the existence of a continuous orientation gradient between the paired variants is confirmed by an EBSD map acquired on a magnesium single crystal. The sample was compressed at 100°C by 5% along its **c**-axis. This test was initially performed to form compression twins; however, the EBSD maps on a cross-section clearly show the formation of extension twins (Fig. 10a). In a first stage, as the result was considered with skepticism; however, in a second stage, we realized that the formation of extension twins in compression tests was already reported and assumed to be due to the unloading step [41][42]. It is indeed possible that during unloading the elastic back-force acts as a tensile force for the specimen. As the unloading mechanical test is highly symmetric, we decided to look again at the EBSD results in the aim to check whether or not the expectations described in the previous section are verified.

The EBSD results are in very good agreement with the theoretical expectations. The extension twins shown in Fig. 10a have a ladder shape, with two parallel long directions along the traces let by the $(0\bar{1}12)$ habit planes, and a series of segments forming the "rungs" of the ladder parallel to the traces let by the $(01\bar{1}2)$ habit planes. A series of complex disconnections movements can probably be imagined in order to explain the different parts of this structure, but we think more probable that this ladder structure was formed suddenly (displacively) "in one breath". The $(0\bar{1}12)$ and $(01\bar{1}2)$ twins differ only by their obliquity angle; they correspond to the twins named ($t_1$, $t_5$) in section 5, and ($D_z^-$, $D_z^+$) in section 6.2. The continuous features expected by the model are experimentally confirmed. Indeed, a circular arc between the **c**-axes of the two variants $D_z^-$ and $D_z^+$ appears in the <001> pole figure (Fig. 10b). The middle of this arc corresponds to the (90°, **a**) variant $D_z$. Rainbow colored images of the orientation gradients in the range [0,10°] are shown for the twins and for the surrounding parent matrix in Fig. 11a and Fig. 11b, respectively. Particularly, Fig. 11a proves that there is not strict grain boundaries between the two variants $D_z^-$ and $D_z^+$ ; the misorientation between them is continuous as confirmed by the profile of Fig. 11c. The rotation axis of the low-misorientations is the **a**-axis (Fig. 11c), as expected by the model. The continuity between $D_z^-$ and $D_z^+$ is explained by the back-stresses generated by the surrounding matrix. Inversely, the lattice distortions create important plastic deformations into the surrounding matrix, which explains the color gradients observed in Fig. 11b.

Fig. 11b shows that the effect of twinning transformation is not localized at the sole parent/twin interface, and thus that the sole structure of disconnections at the interface is not sufficient to explain the twinning mechanism (see section 1.1).

### 6.4. Prediction of the formation of extension twins for negative Schmid factors.

The calculations showed that there is a volume variation during extension twinning (Fig. 3). Since the distance OV also changes (Fig. 6b), the twinning plane can't be let fully invariant during the transformation. It is true that when the transformation finishes, i) the volume comes back to its initial value, ii) the twinning plane is fully restored, and iii) the distortion matrix becomes a simple shear matrix; however, the details of the mechanism between the initial and final state shouldn't be ignored; twinning is not shearing. What is the consequence? The Schmid factor used to calculate the resolved shear stress on the twinning plane is not adapted anymore to predict the twin formation. Is it possible to substitute the Schmid's law by another one? Could it explain some of the abnormalities observed in the formation of the extension twins (see section 1.2)?

It is reasonable to think that the volume change observed during the lattice distortion creates an energy barrier that should be overcome to form the twins. This idea was also suggested for the fcc-bcc martensite transformations [23]. If this assumption is correct, it would mean that instead of using the matrix of complete transformation, i.e. the simple shear matrix (29), one should use the intermediate matrix corresponding to the maximum volume change, i.e. the matrix given by equation (27) with $\kappa = \kappa_i$. Since this matrix is not a simple shear matrix a criterion substituting the Schmid's law should be found. We propose to come back to the general formula giving the interaction work W of a unit volume of a material that deforms, by mechanical twinning or phase transformation, inside an external stress field **Γ**. The scalar W is given by the Frobenius inner product

$$W = \mathbf{\Gamma}_{ij} . \mathcal{E}_{ij} \tag{42}$$

which is simply the integral form of the infinitesimal work $dW = \Gamma_{ij} . d\mathcal{E}_{ij}$ for uncorrelated fields $\mathbf{\Gamma}$ and $\mathcal{E}$, sometimes called "virtual work" in the continuum mechanics textbooks. The work W is performed by the external stress during the transformation; it is used to deform the surrounding environment in which the twin forms. A high value of interaction work means a high probability of transformation, and negative value should correspond to an impossibility of transformation. The interaction energy is the opposite of W; it should be added to the usual form of energy that is calculated to predict a phase transformation. Maximizing the interaction work is equivalent to minimizing the interaction energy. As proved in Appendix A, the interaction work is proportional to the Schmid factor in the case of a simple shear and to the Patel and Cohen criterion [56] in the case of invariant plane strain deformation.

Let us illustrate how equation (42) can be used for extension twinning. The numerical value of the final distortion matrix is given in the basis $\mathbf{B}_{ortho}$ by equation (29):

$$\mathbf{D}_f^{p \to t} = \mathbf{D}_{ortho}^{p \to t}(1/3) \approx \begin{pmatrix} 1 & 0 & 0 \\ 0 & 0.9412 & 0.0624 \\ 0 & -0.0555 & 1.0588 \end{pmatrix} \tag{43}$$

Now, we consider a mechanical test on an hcp crystal that was rotated by an angle $\theta$ around the direction **n** normal to $(0\bar{1}12)$ plane, and tilted by an angle $\phi$ around the **x**-axis. The rotation matrices axis are noted $\mathbf{R}_n(\theta)$ and $\mathbf{R}_x(\phi)$, respectively. As showed in the equation [A4] of Appendix A, the distortion matrix of the rotated-tilted parent crystal, expressed in the orthonormal basis $(\mathbf{x}, \mathbf{y}, \mathbf{z})$ linked to the mechanical test becomes

$$\mathbf{D}_f^{p \to t}(\phi, \theta) = \mathbf{R}_x(\phi) \, \mathbf{R}_n(\theta) . \mathbf{D}_f^{p \to t} . \left(\mathbf{R}_x(\phi) \, \mathbf{R}_n(\theta)\right)^{-1} \tag{44}$$

The twinning deformation matrix is $\mathcal{E}_f^{p \to t}(\phi, \theta) = \mathbf{D}_f^{p \to t}(\phi, \theta) - \mathbf{I}$. According to equation (42), its interaction work with a stress field $\mathbf{\Gamma}$ is

$$W_f(\phi, \theta) = \mathbf{\Gamma}_{ij} . \left(\mathcal{E}_f^{p \to t}\right)_{ij}(\phi, \theta) \tag{45}$$

The graph of interaction work $W_f(\phi, \theta)$ in the case of a tensile test along the **z**-axis is given in Fig. 12a. As expected, it is similar to the graph obtained for a simple shear (Fig. A3); the only difference is that the interaction work is maximum for a shear plane tilted at $\phi = 45°$ in the case of Fig. A3, whereas it is maximum for a crystal tilted at $\phi = 2°$ for extension twinning. This is due to the fact that the $(0\bar{1}12)$ shear plane is already tilted 43° far from the basal plane in a crystal positioned horizontally, which means that a tilt of $\phi_s = 2°$ is sufficient to place the shear plane at 45° from the **z**-axis. According to Fig. 12a the extension twins should form only for positive interaction work, i.e. for tilt angles $\phi_s - \frac{\pi}{4} < \phi < \phi_s + \frac{\pi}{4}$, i.e. in the range of tilt angle $\phi \in [-43°, 47°]$. What happens if the intermediate distortion matrix is used in place of the final (shear) matrix? The numerical values of the intermediate distortion matrix corresponding to maximal volume change is given in the basis $\mathbf{B}_{ortho}$ by equation (27) with $\kappa = \kappa_i$:

$$\mathbf{D}_i^{p \to t} = \mathbf{D}_{ortho}^{p \to t}(\kappa_i) \approx \begin{pmatrix} 1 & 0 & 0 \\ 0 & 0.9852 & 0.0308 \\ 0 & -0.0290 & 1.0449 \end{pmatrix} \tag{46}$$

This matrix is used to calculate the intermediate distortion matrix $\mathbf{D}_i^{p\rightarrow t}(\phi,\theta)$ in a parent crystal tilted around the **x**-axis by an angle $\phi$, and rotated around the **n**-axis by an angle $\theta$, as in equation (44). The intermediate deformation matrix is $\mathcal{E}_i^{p\rightarrow t}(\phi,\theta) = \mathbf{D}_i^{p\rightarrow t}(\phi,\theta) - \mathbf{I}$. This matrix, instead of $\mathcal{E}_f^{p\rightarrow t}(\phi,\theta)$, is now used in equation (45) to calculate the interaction work $W_i(\phi,\theta)$ at the intermediate state. The graph of $W_i(\phi,\theta)$ is given in Fig. 12b. It is clear that the region of positive work $W_i$ is extended in comparison with $W_f$. For example, for $\theta = 0$, which means that the shear direction is not rotated in the twinning plane and is kept at its maximum value (only the parent crystal is tilted), the interaction work $W_i$ is positive in the range of tilt angles $\phi \in [-59°, 59°]$, whereas $W_f$ is positive only in the range $[-43°, 47°]$. In other words, the use of the intermediate distortion matrix allows the predictions of formation of extension twins for some domains of orientations where the classical Schmid factor is negative. The domains $(\phi,\theta)$ where $W_i \geq 0$ and $W_f < 0$ are shown in the graph of Fig. 12c. It gives the orientations where our model predicts extension twinning whereas classical shear theory does not. Since the experiments are often realized by tilting the parent crystal around the **z**-axis instead of the **n**-axis (because **n** depends on which of the six twinning variants is formed), we have plotted in the *Supplementary Material 6* the same graphs as those of Fig. 12d, but calculated by substituting the rotation $\mathbf{R}_z(\theta)$ by the rotation $\mathbf{R}_n(\theta)$.

The range $\phi \in [-59°, 59°]$ predicted by our model is exactly the range of formation of extension twins deduced by Čapek *et al.* [32] after extrapolating a series of neutron diffraction measurements on polycrystalline magnesium samples deformed by tensile tests at different strains (see Fig.6a of Ref. [32]). The agreement with some experimental data is promising. However, besides this encouraging sign, we should admit that there are many experiments reported in literature that we can't yet explain. We have calculated the interaction work for some stress fields implying compression components, but the graphs giving the values $(\phi,\theta)$ where $W_i \geq 0$ and $W_f < 0$ were reduced to zero, which means that our model can't do better than the classical theory for such conditions.

## 7. Conclusion

The model presented in the paper is based on a diplacive paradigm. Deformation twinning is considered as a martensitic transformation; the atoms move collectively by a homogenous lattice distortion. The work presented here differs, however, from the usual "shear" theory. As in our previous crystallographic studies on martensitic transformations [23][24] that showed that the shear matrices should be replaced by angular-distortive matrices to evaluate the lattice distortion *during* the transformation and take into account the size of the atoms, a similar approach was applied to treat the case of the $\{10\bar{1}2\}$ extension twins in hcp metals. Instead of determining the simple shear matrices that restore the lattice and then guessing the trajectories (shuffles) of the atoms that are not at the lattice nodes, we first calculate the atom trajectories assuming that they are hard spheres of constant size, and then we calculate the analytical expression of the lattice distortion. The advantage is that it is sure that the atoms do not interpenetrate each other, i.e. the energy barrier remains reasonable along the transformation path. During the distortion, the unit volume increases up to 3% and the length of the "diagonal" direction in the twinning plane (direction OV) increases up to 1.4%, both before coming back to their initial values when the transformation is complete. These values are too high to be accommodated fully elastically by relaxing the hard-sphere assumption.

The $\{10\bar{1}2\}$ twinning plane is not fully invariant during the distortion; it is just untilted and fully restored when the distortion is finished.

The calculations also show that the stretch distortion deduced by Liu et al.[35] from their observations of (90°, **a**) twinning domains in a submicron-sized samples, is a component of the distortion matrix associated with the usual (86°, **a**) extension twins in bulk samples. The atom trajectories are very similar in the two cases and differ only by a continuous rotation angle varying between 0 and 3.4°. It can be concluded that the Liu et al.'s observations are in agreement with the displacive nature of mechanical twinning, and there is reason to consider them as a sign for a reconstructive mechanism. An analogy with ferroelectrics makes us think that the stretch distortion associated with the (90°, **a**) domains could be a "spontaneous" distortion that is free to occur in nano-objects, and that the usual distortion associated with the (86°, **a**) twins could be a "constrained" one that occurs in bulk samples. The 3.4° compensating angle can then be viewed as an obliquity angle. From the lattice distortion model and analogies with martensitic transformations, an orientation continuity is expected between the two low-misorientated (86°, **a**) twin variants, for example between the $(0\bar{1}12)$ and $(01\bar{1}2)$ variants. This expectation was fully confirmed by EBSD maps acquired on a magnesium single crystal compressed along its **c**-axis (the extension twins are supposed to be formed during the unloading stage).

As extension twinning differs from a shear mechanism in its details, a criterion that generalizes Schmid's law was used. It is based on the interaction work between an external stress field and the deformation field generated by the twinning distortion. The "virtual work" criterion allows the prediction of extension twin formation for parent crystal orientations associated with negative Schmid factors. More precisely, the model predicts the formation of extension twins during uniaxial tensile tests with parent crystals tilted in the range $[-59°, 59°]$ whereas the usual model predicts extension twins only in the range $[-43°, 47°]$. The fact that a critical tilt angle of 59° was already experimentally reported in the literature is very encouraging. The deviation of more than 12° in the predictions made by the two models a) the classical Schmid's law associated with the shear matrix and b) the virtual work criterion associated with the intermediate distortion matrix, proves that it should be experimentally possible to distinguish the right one. The calculations also show that small obliquity angles can have a huge impact on the habit planes and are worth being taken into account into the predictions.

The model in the present state is macroscopic; it is not able to give details on how the distortion is microscopically accommodated. The arrays of dislocations induced by the distortion at the interface (disconnections) and in the surrounding matrix (disclinations) are not yet predicted. The exact nature of these dislocations, the way they are distributed, and their effect in the formation of basal/prismatic ledges at the interface will be the subject of future investigations; we have no doubt that disconnection theory will then play a key role in the calculations.

**Note:**

The second paper treating the case of the $\{10\bar{1}1\}$ contraction twins with the same approach is available on Arxiv [57]. Other papers are also in preparation for other twin modes in hcp metals.

# Acknowledgments

I would like to show my gratitude to Prof. Roland Logé, director of LMTM, for his piece of advice to improve the paper quality, to Mathijs van der Meer for the Gleeble tests, and to PX group for the laboratory subsidy and for our scientific and technical exchanges.

# Appendix A

The aim of this appendix is to show that the Schmid factor for a simple shear deformation is a special case of a more general criterion based on the interaction work between an external stress field and an internal strain field.

The shear matrix of amplitude **s** along the [100] direction on the (001) shear plane is

$$\mathbf{S} = \begin{pmatrix} 1 & 0 & s \\ 0 & 1 & 0 \\ 0 & 0 & 1 \end{pmatrix} \quad [A1]$$

This situation is illustrated in Fig. A1a. Now, we consider that the case where the same shear occurs in a crystal that is rotated around **z** by an angle $\theta$ and tilted around **x** by an angle $\phi$, as shown in Fig. A1b. The two rotation matrices are noted $\mathbf{R}_x(\phi)$ and $\mathbf{R}_z(\theta)$ respectively, and illustrated in Fig. A2. They are

$$\mathbf{R}_x(\phi) = \begin{pmatrix} 1 & 0 & 0 \\ 0 & Cos\phi & -Sin\phi \\ 0 & Sin\phi & Cos\phi \end{pmatrix} \text{ and } \mathbf{R}_z(\theta) = \begin{pmatrix} Cos\theta & -Sin\theta & 0 \\ Sin\theta & Cos\theta & 0 \\ 0 & 0 & 1 \end{pmatrix} \quad [A2]$$

One way to deduce the new shear matrix is to write it from the new coordinates of the shear plane normal **n** and shear vector **s** given by

$$\mathbf{n}(\phi) = \mathbf{R}_x(\phi).[001] = \begin{bmatrix} 0 \\ -Sin\phi \\ Cos\phi \end{bmatrix} \quad [A3]$$

$$\mathbf{s}(\theta) = \mathbf{R}_x(\phi).\mathbf{R}_z(\theta).[010] = \begin{bmatrix} -Sin\theta \\ Cos\phi.Cos\theta \\ Sin\phi.Cos\theta \end{bmatrix}$$

The other way is to write directly the shear (active) matrix **S** in the rotated crystal by:

$$\mathbf{S}(\phi, \theta) = \mathbf{R}_x(\phi)\,\mathbf{R}_z(\theta).\mathbf{S}.\big(\mathbf{R}_x(\phi)\,\mathbf{R}_z(\theta)\big)^{-1} \quad [A4]$$

The two methods gives the same result for the shear matrix [A1] :

$$\mathbf{S}(\phi, \theta) = \begin{pmatrix} 1 & Sin\theta.Sin\phi.s & -Cos\phi.Sin\theta.s \\ 0 & 1-Cos\theta.Cos\phi.Sin\phi.s & Cos\theta.Cos^2\phi.s \\ 0 & -Cos\theta.Sin^2\phi.s & 1+Cos\theta.Cos\phi.Sin\phi.s \end{pmatrix} \quad [A5]$$

The shear displacements are given by the matrix $\mathbf{S}(\phi, \theta) - \mathbf{I}$. When they occur inside a stress field $\mathbf{\Gamma}$, a work is performed; it is the product of force along the displacements. When the strain field and the stress field are not correlated, this work is

$$W(\phi, \theta) = \Gamma_{ij}.\mathcal{E}_{ij}(\phi, \theta) \quad [A6]$$

with the Einstein convention on the repetition of the indices (*i,j*).

Let us consider the simple case of a uniaxial stress $\sigma_z$ applied along the z direction, i.e.

$$\mathbf{\Gamma} = \begin{pmatrix} 0 & 0 & 0 \\ 0 & 0 & 0 \\ 0 & 0 & \sigma_z \end{pmatrix} \qquad [A7]$$

The interaction work becomes

$$W(\phi, \theta) = \sigma_z . (\text{Cos}\theta . \text{Cos}\phi . \text{Sin}\phi) . s = \sigma_z\, m\, s = \tau\, s \qquad [A8]$$

where $\tau$ is the shear stress and $m$ is Schmid factor, usually given by form $m = \text{Cos}\phi . \text{Cos}\lambda$ with $\lambda$ the angle between the direction **z** and the shear vector **s**. From the expression [A3] it is easy to see that $\text{Cos}\lambda = \text{Sin}\phi . \text{Cos}\theta$, and that, indeed, $m = \text{Cos}\phi . \text{Sin}\phi . \text{Cos}\theta$. The advantage of the expression $m = \text{Cos}\phi . \text{Sin}\phi . \text{Cos}\theta$ in comparison to the classical form $m = \text{Cos}\phi . \text{Cos}\lambda$ is that the two angles $\theta$ and $\phi$ are not correlated, whereas the angle $\lambda$ is a function of the angle $\phi$. For instance, it could be thought that it is possible to maximize $m$ by choosing $\phi = 0$ and $\lambda = 0$, but this is not true due to the correlation of the two angles. With the expression $m = \text{Cos}\theta . \text{Cos}\phi . \text{Sin}\phi$, it is easy to see that $m$ is maximized for $\phi = \frac{\pi}{4}$ and $\theta = 0$.

The interaction work [A8] is simply the work performed by the shear stress along the shear displacement. The interaction energy is the opposite value of the interaction work. It is an energy by unit of volume and is expressed, as $\sigma_z$, in MPa (for the units remember that Pa = J/m$^3$). The Schmid factor, used to introduce a critical resolved shear stress, is equivalent to a critical value of the interaction work. As for the Schmid factor, one can plot the interaction work depending on the orientation of the crystal. The graph of $W(\phi, \theta)$ is given in Fig. A3. It can be checked that the interaction work is maximum at $(\phi, \theta) = \left(\frac{\pi}{4}, 0\right)$ and $(\phi, \theta) = \left(-\frac{\pi}{4}, \pi\right)$. It is important to note that this is just a special case of the equation [A6] . The big advantage of equation [A6] lays in its generality; it doesn't depend on the type of distortion and can be applied to non-shear distortions. For example, the well-known Patel-Cohen equation [56] that gives the interaction energy for an invariant plain strain (simple shear + deformation perpendicularly to the habit plane) is also a special case of equation [A6] .

# Figures

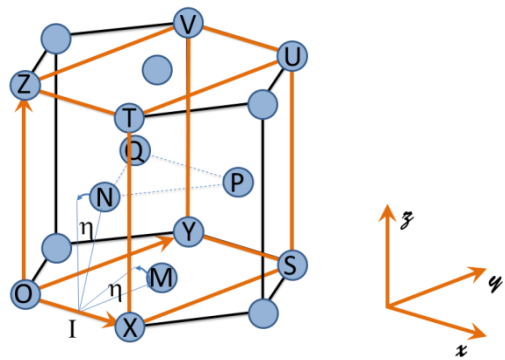

*Fig. 1. Hexagonal lattice of the parent crystal with its associated orthonormal basis (**x**,**y**,**z**) , **x** = **a**$_p$, **y** = 2**a**$_p$ + **b**$_p$, **z** = **c**$_p$. Some positions of the Mg atoms are labeled in order to describe the atomic displacement during the extension twinning process. The angle $\eta$ is used as the unique parameter of the lattice distortion.*

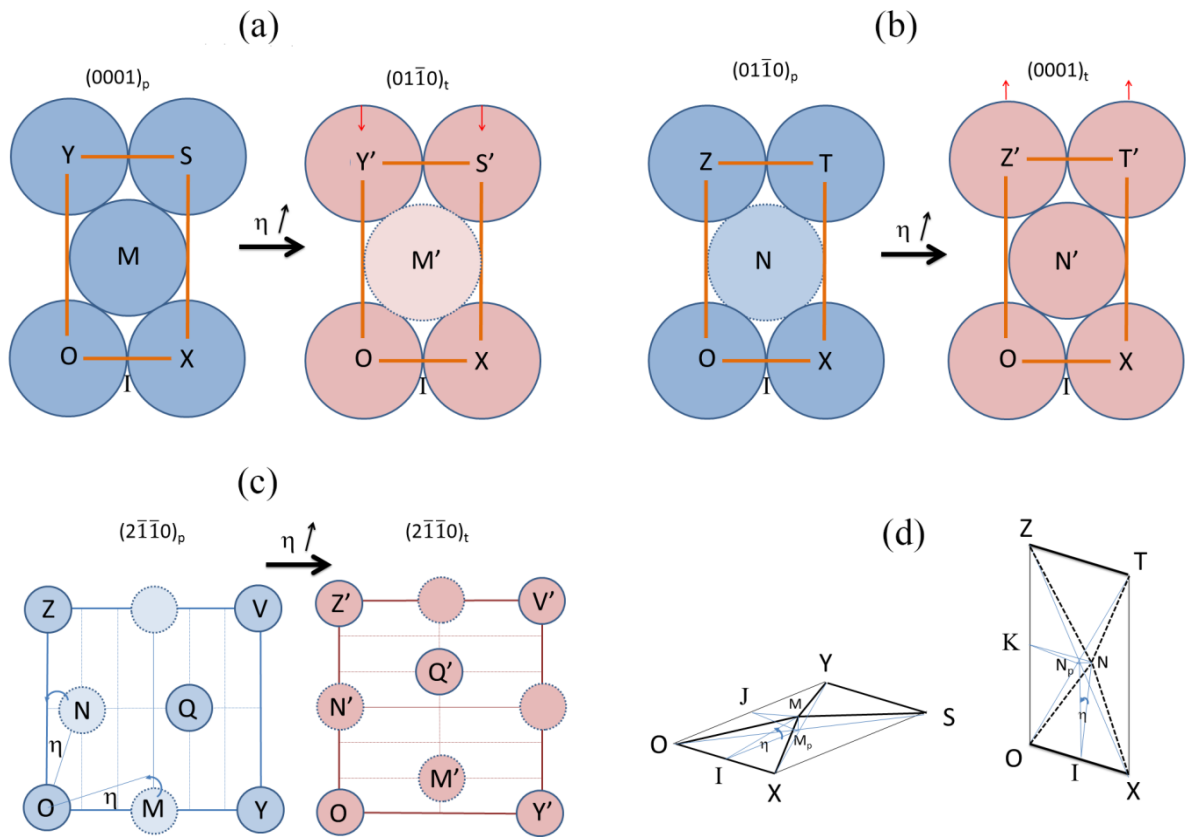

*Fig. 2. Schematic representation of the atomic displacements and lattice distortion during extension twinning viewed on different planes of the parent crystal: a) on OXY = $(0001)_p$, b) on OXZ = $(01\bar{1}0)_p$, c) on OYZ = $(2\bar{1}\bar{1}0)_p$ planes. The parent crystal is in blue and its resulting extension twin is in salmon. All the atomic displacements are functions of a unique parameter, i.e. the angle $\eta$ of rotation of the M atom around the **OX** axis. The vector **OX** = **a** remains invariant. (d) Schemes explaining the change of the distances OY and OZ: as the atom M moves far from plane (OXY), the distance OY decreases, and as the atom N moves toward the plane (OXZ), the distance OZ, initially OZ = c, increases to eventually become OZ' = a when the distortion is complete. The $(0001)_p$ and $(01\bar{1}0)_p$ planes are transformed into the $(01\bar{1}0)_t$ and $(0001)_t$ planes, respectively. During the distortion process, the hard-sphere packing assumption imposes that MO = MX = MY = MS = NO = NX = NT = NT = OX = atomic diameter.*

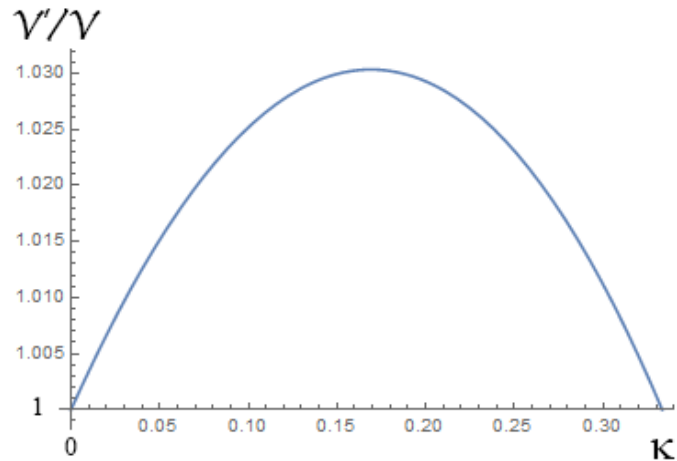

*Fig. 3. Change of volume ratio $\mathcal{V}'/\mathcal{V}$ during extension twinning, as function of the parameter $\kappa = Sin(\eta)$, varying from $\kappa_s = 0$ (start) to $\kappa_f = 1/3$ (finish).*

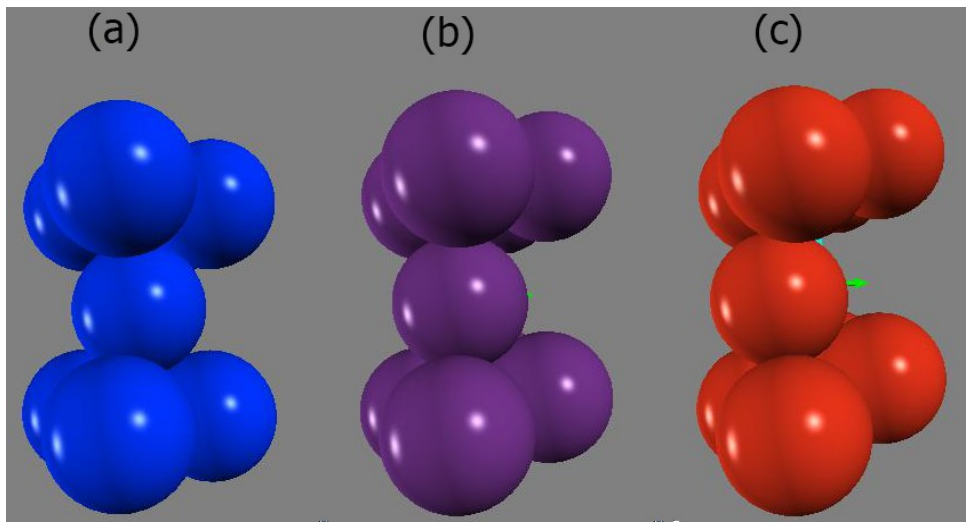

*Fig. 4. 3D view of the stretch distortion of a hexagonal Bravais unit cell. (a) Initial hcp cell ($\eta = 0$), with the $(0001)_p$ plane horizontal and $(01\bar{1}0)_p$ plane vertical, (b) intermediate state ($\eta = 9°$), and (c) final state ($\eta \approx 20°$). The final state c) is a restored hcp structure, but with the basal and prismatic planes interchanged in comparison with (a).*

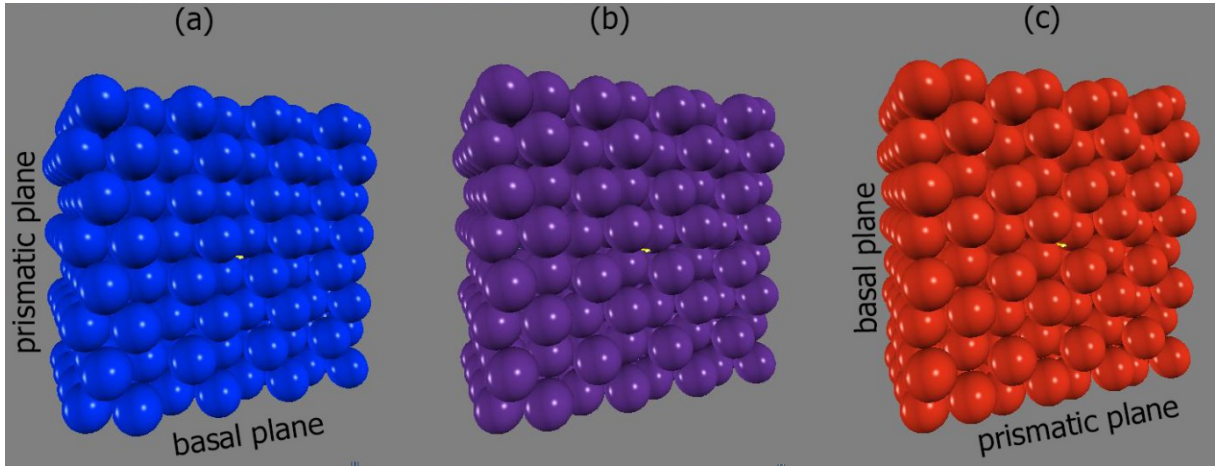

*Fig. 5. 3D view of the stretch distortion of a crystal made with 4x4x4 XYZ cells. (a) Initial hcp cell ($\eta = 0$), (b) intermediate state ($\eta = 9°$), and (c) final state ($\eta \approx 20°$). The basal and prismatic planes interchanged during the distortion.*

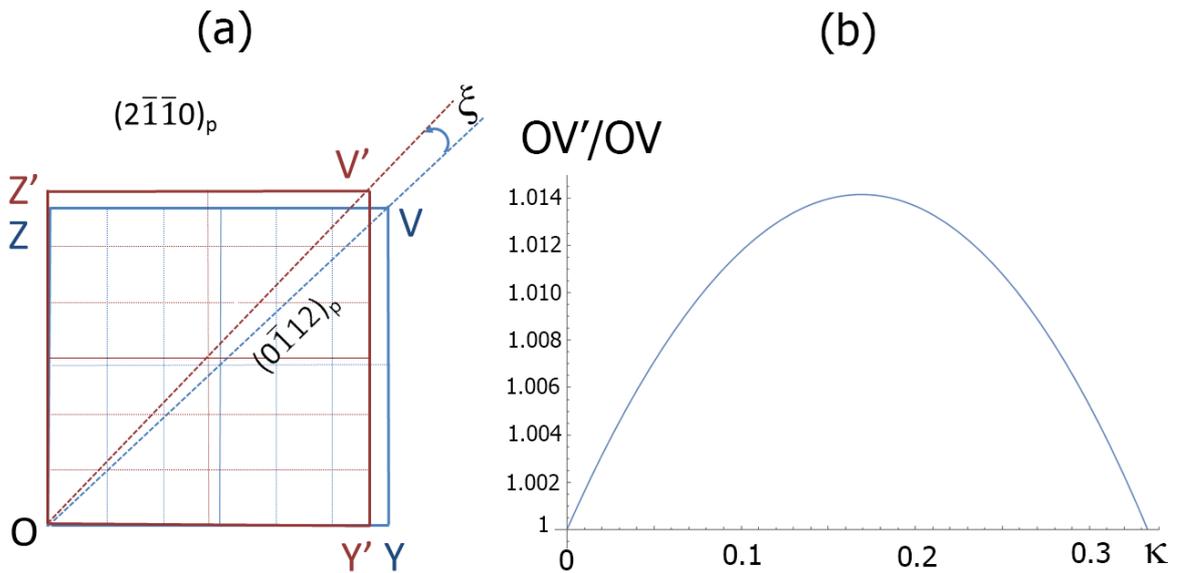

*Fig. 6. Change of the direction **OV** during twinning distortion. (a) Schematic view on the plane OYZ = $(2\bar{1}\bar{1}0)_p$ of the tilt $\xi$ of **OV** around the **a**-axis. (b) Evolution of the ratio of distances OV'/OV proving that even if the tilt $\xi$ is corrected, the $(0\bar{1}12)$ plane cannot be maintained fully invariant.*

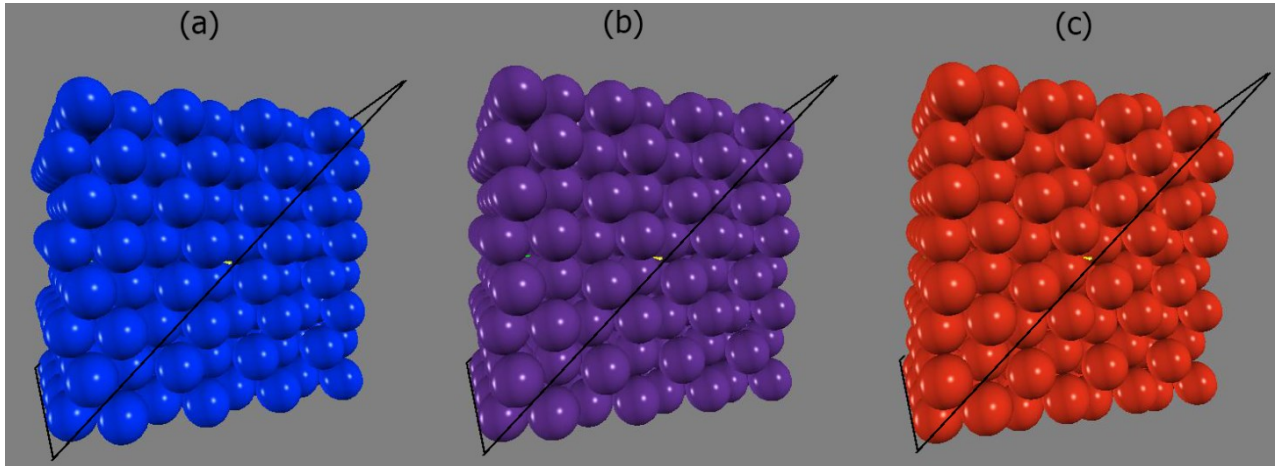

*Fig. 7. 3D view of the distortion associated with $\{10\bar{1}2\}$ extension twinning with a crystal made with 4x4x4 XYZ cells. (a) Initial hcp cell ($\eta$ = 0), (b) intermediate state ($\eta$ = 9°), and (c) final state ($\eta \approx 20°$). The hcp structure is restored but the basal and prismatic planes interchanged while the $(0\bar{1}12)$ plane is maintained untilted. The $(0\bar{1}12)$ plane is marked by the black-lined section.*

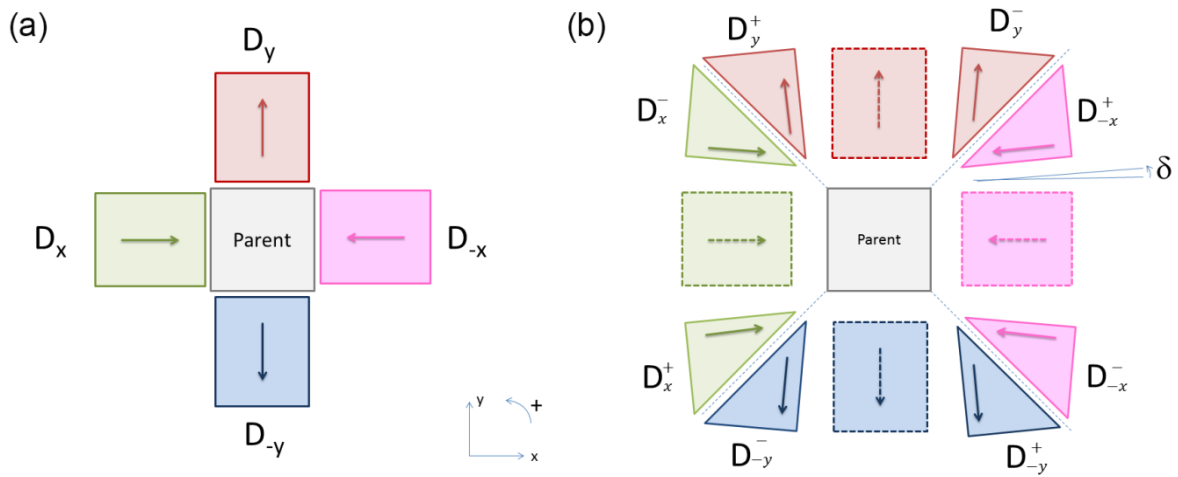

*Fig. 8. 2D representation of the orientational domains in a ferroelectric crystal created by a cubic→ tetragonal phase transition. (a) Domains of "spontaneous" distortion with the polarization vectors (arrows) along the high-symmetry **x** and **y**-axes. (b) Domains experimentally observed, with domain walls on the (1,1) and (1,-1) planes. A small rotation of the spontaneous domains by an angle δ, called obliquity, is required to respect the compatibility conditions at the interfaces. There are 4 variants in case (a) and 8 in case (b).*

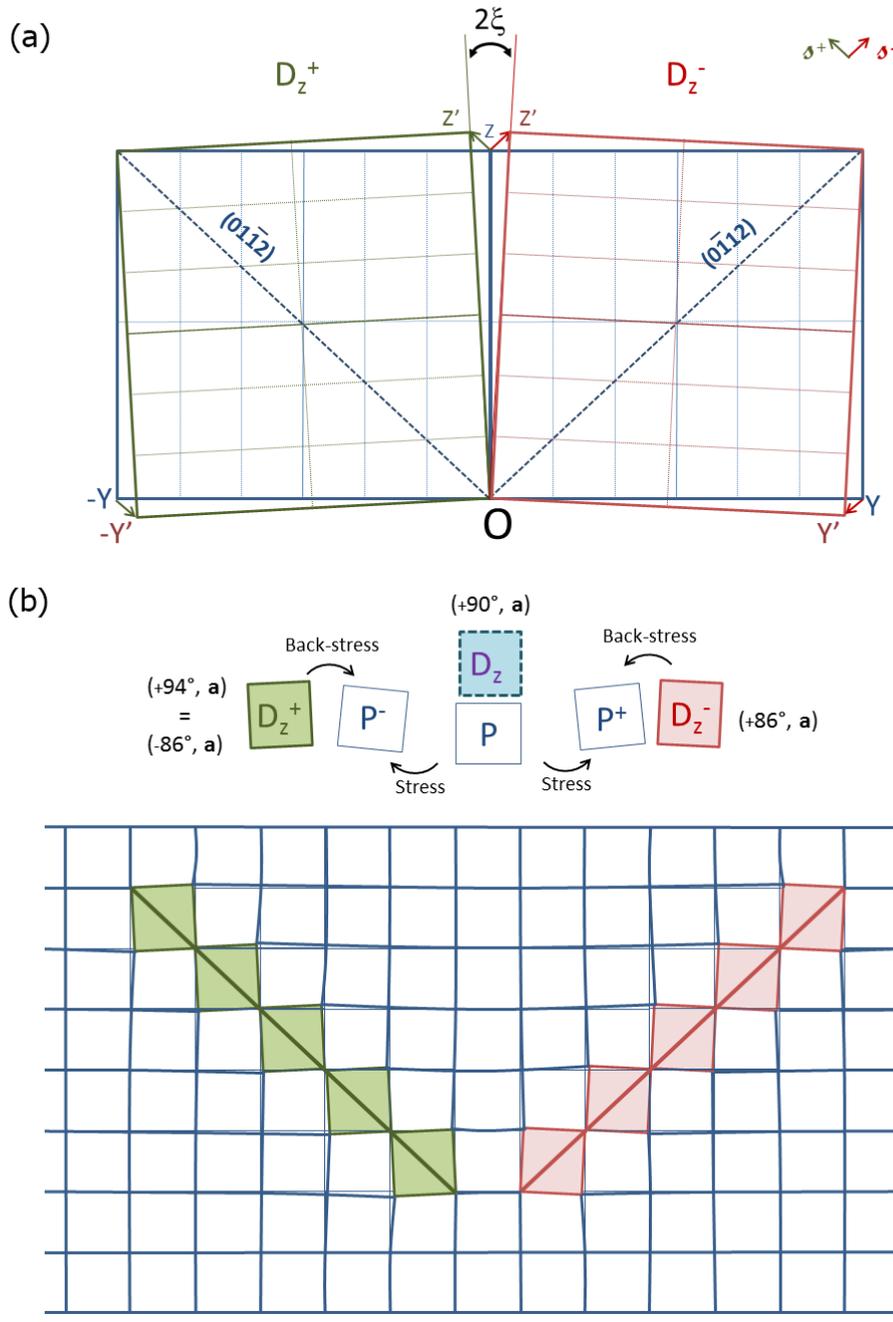

*Fig. 9. Concomitant formation of a pair of low-misoriented (86°, **a**) variants. (a) In analogy with ferroelectrics, the variants $t_1$ and $t_5$ are associated to an obliquity angle of $\xi$ = -3.4° and $\xi$ = 3.4°, and named $D_z^-$ and $D_z^+$, respectively. Their habit plane is $(0\bar{1}12)$ and $(01\bar{1}2)$, respectively. (b) In analogy with previous works on martensitic transformations, it is expected that the back-stresses generated by the dislocations created by the twinning distortion in the surrounding matrix induce a continuous rotation between the two variants $D_z^-$ and $D_z^+$. Both have the tendency to be re-oriented back to the initial lattice orientation (before distortion), which corresponds to the intermediate (90°, **a**) variant $D_z$. A continuous rotation of angle $2\xi$ around the **a**-axis is expected to link the variants $D_z^-$ and $D_z^+$.*

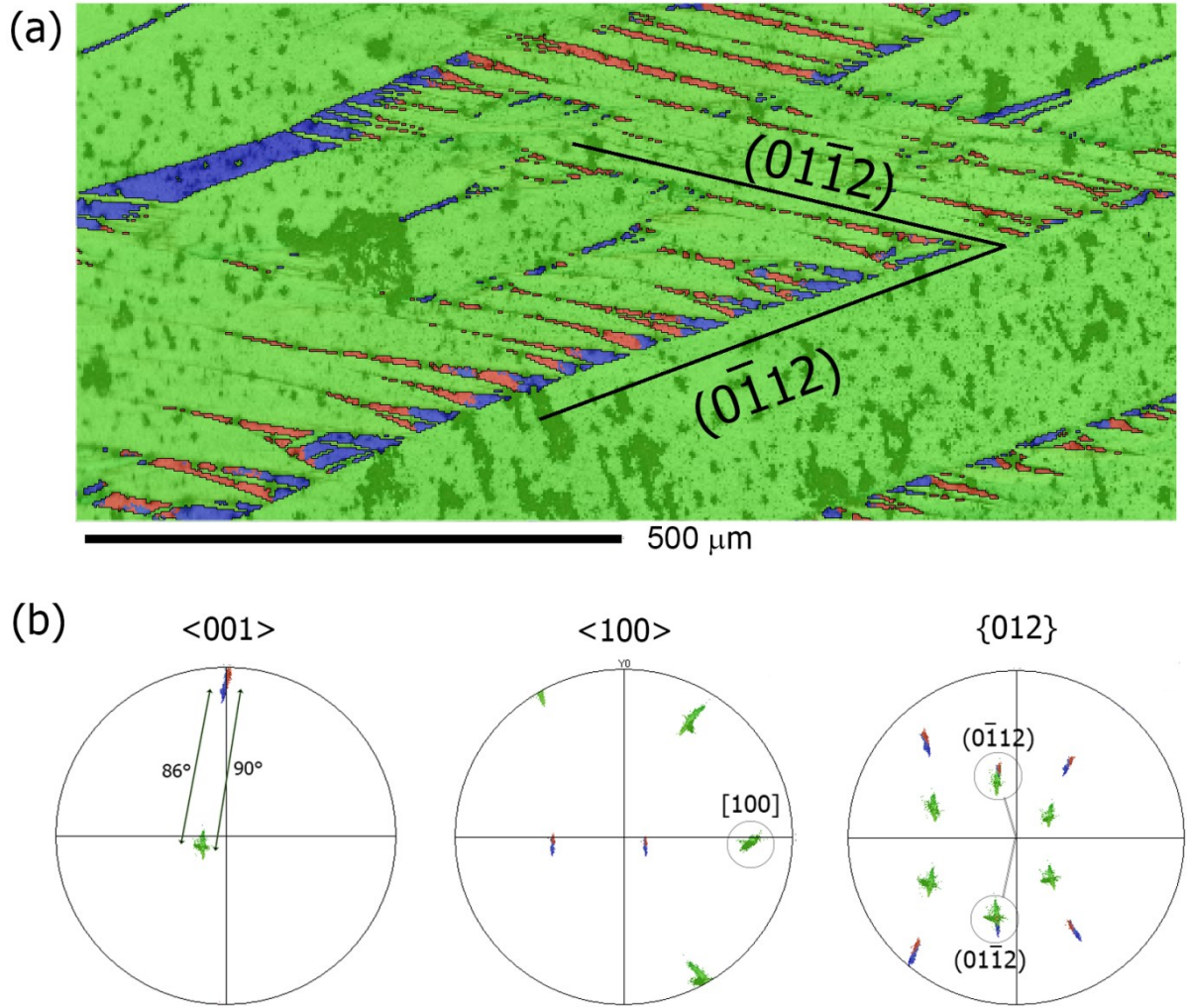

Fig. 10. EBSD map in longitudinal cross-section of a magnesium single-crystal compressed at 100°C along the **c**-axis, probably during the unloading stage. (a) Orientation maps in Euler colors. The parent crystal is in green, the twins are in red and blue. (b) <001>, <100> and {012} pole figures. They show that the red and blue crystals are a pair of low-misoriented (86°,**a**) twin variants. The directions normal to the traces of habit planes and reported in the {012} pole figure show that the habit plane of the blue twins is $(0\bar{1}12)$ and that of the red twin is $(01\bar{1}2)$. The continuous orientation gradient expected from the model is confirmed by the circular red-blue arc around the common **a**-axis encircled in the <100> pole figure. The middle of the arc corresponds to the (90°, **a**) twin.

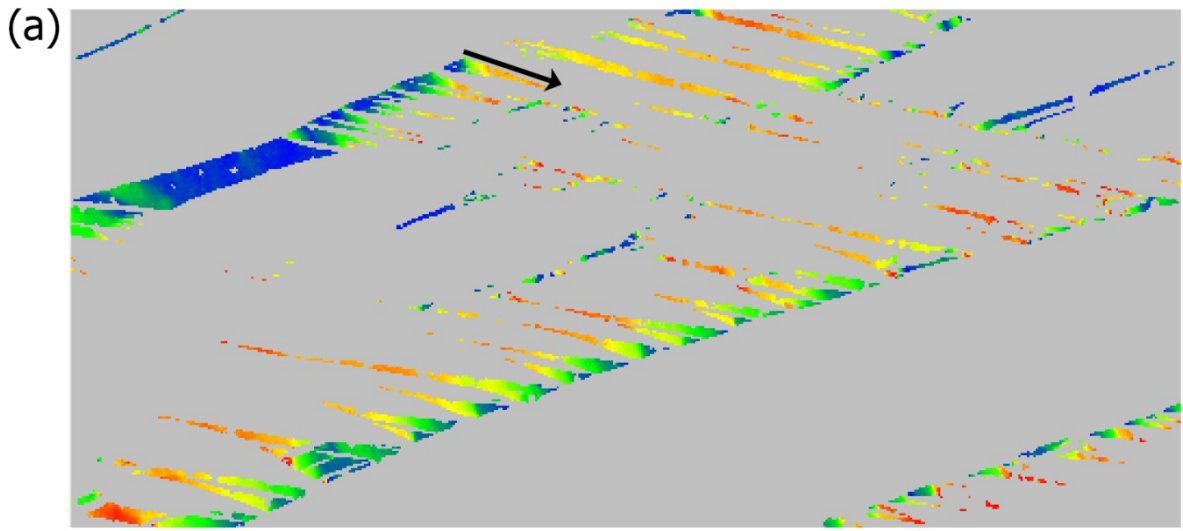

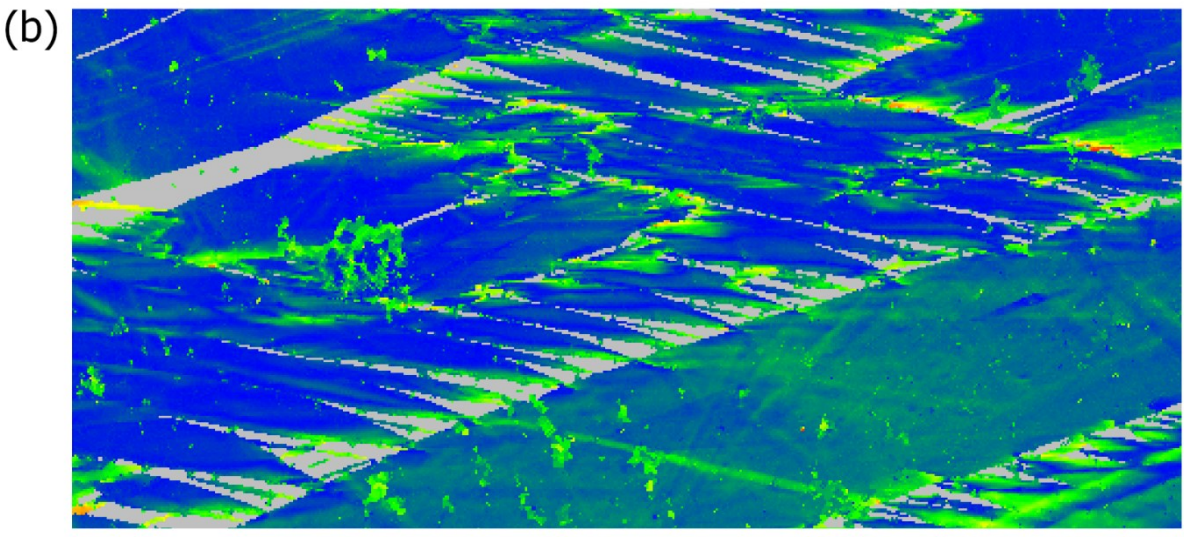

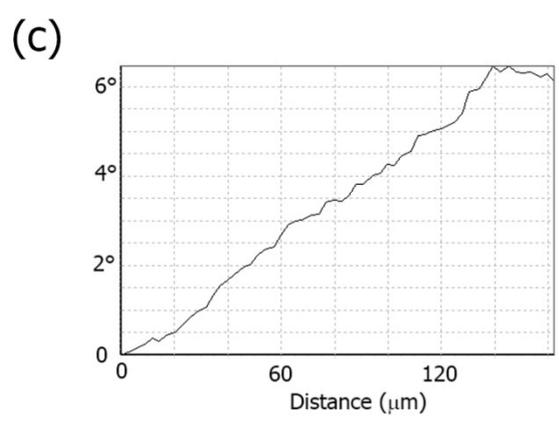
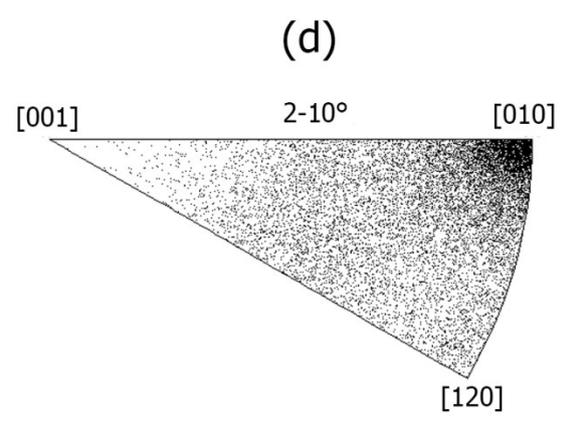

*Fig. 11. EBSD rainbow coloring of the continuous orientation gradients in the range [0,10°]. (a) Orientation gradients in the extension twins, (b) orientation gradients in the surrounding parent matrix. (c) Misorientation profile along the bold arrow marked in (a). (d) Rotation axes associated with the low-angle misorientations (rotation angle between 2° and 10°) plotted in the inverse pole figure.*

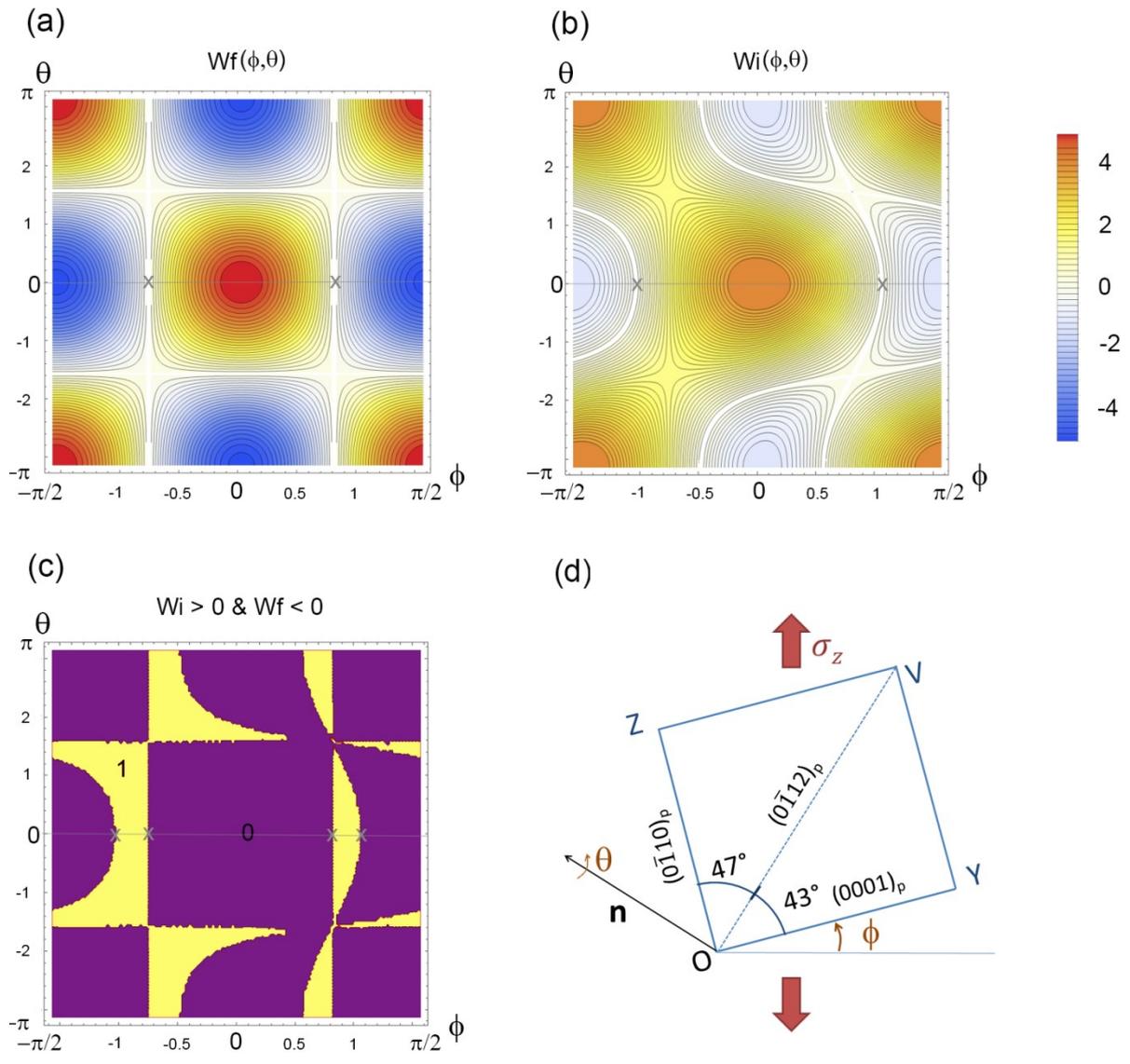

Fig. 12. Interaction work (in MPa) during extension twinning in a tensile stress field oriented along the **z**-axis of a parent crystal tilted by an angle $\phi$ around the **a**-axis and rotated by an angle $\theta$ around the **n**-axis (normal to the twinning plane). a) Interaction work $W_f$ calculated with the complete distortion (shear) matrix. $W_f$ is proportional to the usual Schmid factor. b) Interaction work $W_i$ calculated with the intermediate distortion matrix corresponding to the maximum volume change. c) Graph showing in yellow the orientations $(\phi,\theta)$ where the condition $W_i > 0$ & $W_f < 0$ is true, i.e. where the criterion based on $W_i$ could explain the twin formation despite negative Schmid factors ("anomalous" twins). The axes in the graphs are in radians. d) Schematic view of the orientation of the parent crystal. The segments delineating the domains $W_i > 0$ & $W_f < 0$ for $\theta = 0$ are marked by the grey crosses in a) and b), respectively.

# Appendix Figures

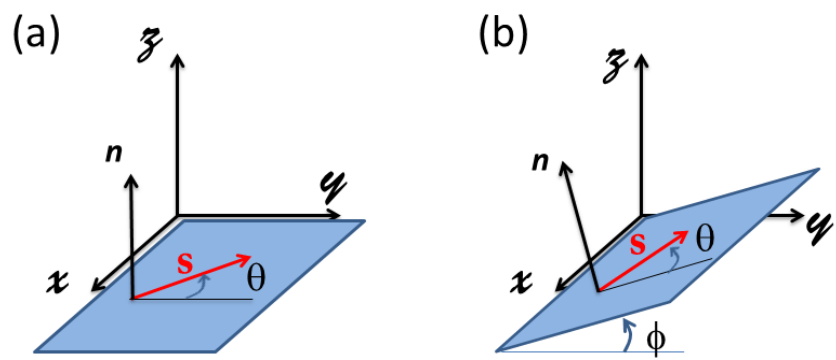

*Fig. A1.  Schematic view of a simple shear with the shear vector **s** that is rotated by an angle $\theta$ in the shear plane of normal **n**. a) The shear plane is horizontal, and b) the shear plane is tilted by an angle $\phi$ around the **x**-axis.*

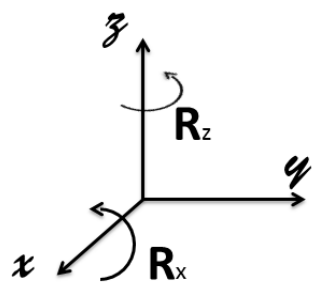

*Fig. A2.  Graphic representation of the two rotations $R_x$ and $R_z$.*

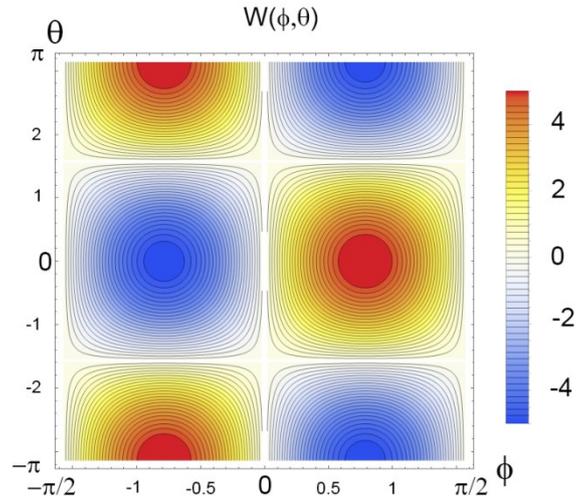

*Fig. A3. Interaction work (in MPa) performed by a simple shear distortion that occurs in a tensile stress field oriented along the **z**-axis with a parent crystal that is tilted by an angle $\phi$ around the **x**-axis and rotated by an angle $\theta$ around the **n**-axis, as illustrated in Fig. A1b. The graph is calculated from equation [A8] with $\sigma_z = 100 \, MPa$ and s = 0.1. The maximum and minimum values are obtained in the red and blue area, respectively; they are $W_{max}$ = 5 MPa and $W_{min}$ = -5 MPa.*